\documentclass[usenatbib,usegraphicx]{mn2e}
\usepackage{url}
\usepackage{times}
\usepackage{aas_macros}
\newcommand{\Rev}[1]{#1}
\newcommand{\D}{\ensuremath{\delta}}
\newcommand{\proj}{\ensuremath{_\mathrm{proj}}}
\newcommand{\app}{\ensuremath{_\mathrm{app}}}
\newcommand{\de}{\ensuremath{\mathrm{d}}}
\newcommand{\dL}{\ensuremath{d_\mathrm{L}}}

\newcommand{\pr}{\ensuremath{^\prime{}}}
\newcommand{\dpr}{\ensuremath{^{\prime\prime}{}}}

\newcommand{\magfac}{\ensuremath{\mathcal{M}}}
\newcommand{\Bmin}{\ensuremath{B_\mathrm{min}}}
\newcommand{\obs}{\ensuremath{_\mathrm{obs}}}
\newcommand{\Lp}{\ensuremath{\Lambda\proj}}
\pubyear{2008}

\title[How relativistic jets look]{Retardation magnification and the
  appearance of relativistic jets\thanks{Accepted by MNRAS 2008 June
    23. Received 2008 June 23; in original form 2008 February 23.}}

\author[S.~Jester]{Sebastian Jester\thanks{Portions of this work were
    carried at the Particle Astrophysics Center, Fermilab MS 127, PO
    Box 500, Batavia, IL 60510, USA; and while the author was an Otto
    Hahn fellow of the Max-Planck-Gesellschaft at the Department of
    Physics and Astronomy, University of Southampton, Southampton SO17
    1BJ, United Kingdom}\thanks{E-mail:
    jester@mpia.de}\\ Max-Planck-Institut f\"ur Astronomie,
  K\"onigstuhl 17, 69117 Heidelberg, Germany}

\begin{document}

\maketitle

\begin{abstract}
Thanks to the availability of high-resolution high-sensitivity
telescopes such as the Very Large Array, the \emph{Hubble Space
  Telescope}, and the \emph{Chandra X-ray Observatory}, there is now a
wealth of observational data on relativistic jets from active galactic
nuclei (AGN) as well as galactic sources such as Black-Hole X-ray
Binaries.  Since the jet speeds cannot be constrained well from
observations, but are \Rev{generally} believed to be relativistic, physical
quantities inferred from observables are commonly expressed in terms
of the unknown beaming parameters: the bulk Lorentz factor and the
line-of-sight angle, usually in their combination as relativistic
Doppler factor.  This paper aims to resolve the discrepancies existing
in the literature about such ``de-beaming'' of derived quantities, in
particular regarding the minimum-energy magnetic field estimate.  The
discrepancies arise because the distinction is not normally made
between the case of a fixed source observed with different beaming
parameters and the case where the source projection on the sky is held
fixed. The former is usually considered, but it is the latter that
corresponds to interpreting actual jet observations.  Furthermore,
attention is drawn to the fact that apparent superluminal motion has a
spatial corollary, here called ``retardation magnification'', which
implies that most parts of a relativistic jet that are actually
present in the observer's frame (a ``world map'' in relativity
terminology) are in fact hidden on the observer's image (the ``world
picture'' \Rev{in general, or ``supersnapshot'' in the special case of
astronomy}).
\end{abstract}


\section{Introduction}
\label{s:intro}

For over 50 years from the appearance of the seminal \emph{Zur
  Elektrodynamik bewegter K\"orper} \citep{Einstein05}, only the
Lorentz transformations of the 4-coordinates of ``events'' were
considered in the literature, but not how relativistically moving
bodies would \emph{appear} when looked at or photographed.  This was
first done independently and (relatively) simultaneously by
\citet{Pen59} and \citet{Terrell59}; the former showed that the
projected outline of a relativistically moving sphere is always a
circle, while the latter provided a more extensive discussion of the
appearance of moving bodies and pointed out the key features of
observing relativistically moving bodies: they appear both
\emph{rotated} and \emph{scaled} (more details will be given below).

The motivation for writing the present paper is work on interpreting
observations of relativistic jets \citep[e.g.,]{JHMM06}, where the
need arises to infer physical properties of the jet fluid in its own
rest frame from observations, subject to corrections due to
relativistic beaming, whose magnitude is, however, not known from
observations.  A particular quantity of interest is the rest-frame
minimum-energy magnetic-field estimate for a synchrotron source
\citep{Bur59}, and there are different opinions in the literature
about how the true rest-frame minimum-energy field scales with the
relativistic Doppler factor compared to that inferred assuming a
non-relativistic source \citetext{compare eqn.~A3 of \citealt{SSO03}
  to eqn.~A7 of \citealt{HK02}}.  Some of the argument revolves around
whether the observed morphological features of jets are ``blobs'' or
``jets'', and their (apparently) different beaming properties.

Given these differences of opinion on how to ``de-beam'' properly, it
is perhaps surprising that NASA's Astrophysics Data System lists only
three papers on interpreting jet observations as citing
\citet{Terrell59}, and his results do not seem to be part of the
common knowledge of jet researchers.  One of the citing papers is
\citet{LB85}, who consider the implications of relativistic beaming on
the difference between observed and intrinsic source counts and give
detailed formulae for relating observed and jet-frame fluxes and
emissivities.  Some of these formulae had already been presented in
the seminal paper by \citet{BK79}.

It appears that the difficulties in interpreting jet observations
arise because the problem under consideration is ill-posed. As will be
argued in detail below, what matters for interpreting jet observations
subject to unknown beaming paramters is that we have observed the
2-dimensional projection of a source's appearance onto the plane of
the sky and try to infer the source's rest-frame properties from this
projection.  Confusion arises because most formulae in the literature
consider what happens to the observed quantities when a \emph{fixed
  source} moves with different Lorentz factors and at different
line-of-sight angles to the observer, while in observations, it is the
\emph{projection} of the source which is held constant.
\Rev{Furthermore, the effects of light-travel time delays along the
  line of sight are typically only mentioned explicitly in work
  comparing jet simulations to observations, e.g., in \citet{AMGea03}
  and \citet{SwiftHughes08}, but not in the observational literature.}
This and the preference for adopting the fixed-source view may be
related to the fact that Lorentz transformations are usually covered
in great detail in a typical course on special relativity, but
\citet{Pen59} and \citet{Terrell59} are hardly mentioned in relativity
textbooks.

This leads to the present paper with the following outline: the
remainder of the introduction summarizes the results by \citet{Pen59}
and \citet{Terrell59} and sets out some basic definitions and
terminology.  The appearance of relativistic objects, and in
particular of astrophysical jets, is discussed in
\S\ref{s:appearance}, both from a theoretical point of view and using
a simple ray-tracer. Ready-to-use formulae for relating jet-frame
quantities to observables are given in \S\ref{s:beaming}, including
the minimum-energy field.  The discussion and summary are given in
\S\ref{s:disc}, while Appendix~\ref{s:illustr.lor} describes some
\emph{Gedankenexperimente} on non-conventional world-map measurements
that lead to length expansion and time acceleration.

\subsection{World Pictures and Supersnapshots}
\label{s:intro.worldsnap}

As first noted by \citet{Terrell59}, there is a fundamental difference
in relativity between the \emph{locations} (4-coordinates) of events
as judged by observers that are local to the events and equipped with
sets of clocks that are synchronized in their rest frame, and the
\emph{appearance} of relativistically moving bodies as judged by
distant observers by means of photons that are received
simultaneously; a little earlier, \citet{Pen59} had considered the
special case of the observed outline of a relativistically moving
sphere.  The set of event locations is a \emph{world map}, while the
picture that is taken of the events is a \emph{world picture}.  In the
special case of photons arriving at right angles to the detector
taking the world picture, it is called a \emph{supersnapshot}
\citep{Rindler77}.  Astronomical observations clearly fall under the
definition of a supersnapshot.

The appearance of relativistically moving objects in a supersnapshot
is governed by two aspects of photon paths in special relativity
\citep{Terrell59,Rindler77,LB85}:
\begin{enumerate}
\item \Rev{Two photons traveling abreast with a separation $\Delta s$
  in one frame (i.e., photons traveling ``alongside each other'' with
  $\Delta s$ measured perpendicular to their direction of motion) do so in
  \emph{all} frames. This is the case because $|\Delta s|^2$ is
  invariant under Lorentz transformations and $\Delta s$ is a
  space-like interval.}
\item If a photon is traveling at an angle $\theta\pr$ to the
  direction of motion of some frame that is moving with speed $\beta
  c$ and Lorentz factor $\Gamma = (1-\beta^2)^{1/2}$ with respect to
  an observer, the angle between the direction of motion and the
  photon direction in that frame is related to the angle $\theta$
  between the direction of motion and the photon direction in the
  observer's frame by
\begin{equation}
\mu\pr = \frac{\mu - \beta}{1-\beta\mu},
\label{eq:costrans}
\end{equation}
 where $\mu = \cos\theta$ etc., or, equivalently,
\begin{equation}
\sin \theta\pr = \D \sin\theta,
\label{eq:sintrans}
\end{equation}
where \D\ is the relativistic Doppler factor
\begin{equation}
\D = \left[\Gamma (1-\beta \mu) \right]^{-1}.
\label{eq:Dopplerdef}
\end{equation}
\end{enumerate}
The latter phenomenon is the well-known angle aberration; the former
is perhaps less well-known, but essential for the analysis of images
of relativistically moving objects, and implies that the supersnapshot
is a \emph{scaled} version of the rest-frame image.  Taken together,
they yield Terrell's result that the appearance of such an object in a
supersnapshot is simply the object's appearance as seen from the
aberrated angle $\theta\pr$ in its rest frame, with its apparent size
along the direction of motion scaled by the Doppler factor $\D$.

As a consequence of eq.~(\ref{eq:costrans}), even approaching
objects appear to be seen ``from behind'' unless $\mu < \beta$, i.e.,
$\D > \Gamma$; in the limiting case $\mu=\beta \Leftrightarrow
\D=\Gamma \Leftrightarrow \sin\theta = 1/\Gamma$, a relativistic
object is seen exactly side-on in its rest frame and with exactly its
rest-frame length as its ``projected'' length.

\subsection{Terminology: ``Blobs'' \emph{versus} ``jets''
  \emph{versus} ``shocks''  \Rev{-- at rest in different frames}}
\label{s:intro.terminology}

It is useful clearly to set out the terminology for the remainder of
the paper, because the brightness pattern observed in astrophysical
jets can be at rest in frames that are different from both the
observer frame, and the fluid rest frame, as discussed in detail by
\citet{LB85}.  \Rev{Their discussion and notation is adopted here. It
distinguishes between ``blob'',  ``jet'' and ``shock'' features, which
are defined by being at rest in one of three frames relevant to the
problem.  Thus, it is useful to give the definitions of the relevant
frames together with those of the morphological terms:}
\begin{enumerate}
\item \Rev{The ``observer frame'' is that in which the astronomer is at
  rest.  Once appropriate cosmological corrections are applied, the
  observer frame is conceptually identical to the frame in which the
  jet source and its host are at rest.}

  \Rev{A ``jet'' feature is then a \emph{resolved} brightness pattern whose
  outline is at rest in the observer frame. Observer-frame quantities
  have no primes, e.g. $j$ for volume emissivity.}
\item \Rev{The ``fluid frame'' is the rest frame of the emitting
  fluid, which is taken to be moving through the observer frame at
  relativistic speed.  The term ``rest frame'' is used interchangeably
  with ``fluid frame''.}\footnote{The emitting fluid is not
  necessarily identical with that carrying the bulk of the jet's
  kinetic energy, nor are those two fluids necessarily moving at the
  same speed \protect\citep{HK07}. However, this distinction does not
  affect the relation between observables and physical quantities in
  the rest frame of the emitting fluid, which is the subject of this
  paper. Nevertheless, it needs to be kept in mind when interpreting
  fluid-frame quantities.}

  \Rev{A ``blob'' or ``plasmoid'' is a brightness pattern whose outline
  is at rest in the fluid frame.  Fluid-frame quantities will be
  designated by double primes, e.g. $j\dpr$.}
\item A ``pattern'' or ``shock'' feature is a brightness pattern whose
  outline is at rest neither in the fluid nor in the observer frame,
  \Rev{e.g., a shock traveling through the jet fluid.  It defines the third
  frame, the frame in which this pattern is at rest.}  Pattern-frame
  quantities have single primes, e.g. $j\pr$.  The emissivity of the
  fluid traveling through such a ``shock'' transforms according to the
  fluid's Doppler factor $\D\dpr$, while its projected appearance and
  morphology are governed by the shock's Doppler factor $\D\pr$.
\end{enumerate}
These are fairly intuitive definitions. \Rev{Nevertheless, the difference
between the ``blob'' and ``jet'' formulae in eqn.~C7 of \citet{BBR84}}
is just one of \emph{choice of integration boundaries}, and in
particular whether the integration boundaries are held fixed in the
observer frame when $\D$ is changed (jet case) or are allowed to vary
according to the different projected morphology of a ``blob'' under
changes of $\D$. Thus, it is possible to apply a ``jet'' formula to a
small segment of a blob as long as the integration boundaries are held
fixed in the observer frame.  In \S\S\ref{s:beaming.blobs.restframe}
and \ref{s:beaming.jets.restframe} below, I will present detailed
formulae for converting observed to fluid-frame properties in each
case, with expressions for the minimum-energy field in
\S\ref{s:disc.Bmin.obsfixed}.

\subsection{Basic definitions and beaming formulae}
\label{s:intro.beamformulae}

This section summarizes the basic definitions of surface
brightness/intensity, flux density and luminosity of astronomical
sources, as well as the beaming properties of blobs, jets, and
shocks. I will give explicit formulae for observed surface brightness
and total flux in terms of source parameters for simple geometries, as
well as ray-tracing images showing the appearance of such sources in
supersnapshots.  

For the computation of surface brightness and flux, I use the notation
and formulae as given by \citet{BK79} and \citet{LB85}, assuming an
optically thin, isotropic emission with a power-law emissivity
$j_{\nu} \propto \nu^{\alpha}$ that is constant within the emitting
region.  All observables will be expressed in terms of the emissivity
$j\dpr$ in the fluid rest frame and the source size in the pattern
frame $\Sigma\pr$, which is identical to the fluid and observer frame
for a ``blob'' and ``jet'', respectively.  \Rev{Cosmological
  transformations, however, are not always given explicitly in order
  to simplify the notation; they can be re-incorporated in the usual
  way by inserting appropriate powers of $(1+z)$ for cosmological
  redshifts, and using the appropriate cosmological distance
  measures.}

The surface brightness or intensity $I$, flux density $S_\nu$ and
luminosity $L$ of a source are given by
\begin{eqnarray}
I_\nu &=& \int_0^{s} j_\nu \de x,\\
S_\nu &=& \int_A I_\nu \de A\\
      &=& \dL^{-2} \int_V j_\nu \de V,\\
L_\nu &=& 4 \pi \dL^2 S_\nu \nonumber \\
      &=& 4\pi \int_V j_\nu \de V, \label{eq:Ldef}
\end{eqnarray}
where \dL\ is the luminosity distance to the source, which has
specific emissivity $j$, volume V and projected surface area A.

The transformation properties of these quantities then follow from the
relativistic invariance of $I_\nu/\nu^3$ and the volume transformation
\citep[taken from Appendix C of][]{BBR84}:
\begin{eqnarray}
\nu &=& \D\dpr\; \nu\dpr \label{eq:nutrans} \\
\de\Omega &=& \D\dpr^{-2}\;\de\Omega\dpr\\
I_\nu(\nu) &=& \D\dpr^3\;I\dpr_{\nu\dpr}(\nu\dpr)\\
j_\nu(\nu) &=& \D\dpr^2\;j\dpr_{\nu\dpr}(\nu\dpr) \label{eq:jtrans}
\end{eqnarray}
Assuming optically thin emission makes the discussion appropriate for
arcsecond-scale jets, where sources are not compact enough for
self-absorption to become important.  The difficulties of interpreting
observations of optically thick sources, such as compact cores and
milli-arcsecond scale jets, have been highlighted by \citet{BK79} and
\citet{LB85}. \Rev{The essential point here is that the appearance of
  optically thick sources varies as function of viewing direction, and
  the relativistic angle aberration implies that the beamed appearance
  is governed by this intrinsic viewing angle dependence in addition
  to the flux and surface brigthness beaming.} The volume
transformation deserves separate consideration.

\subsection{Volume transformation of relativistic objects in astronomical images}
\label{s:intro.voltrans}

The fact that the outline of the different kinds of brightness pattern
is at rest in different frames has led some authors to write down
different volume transformation formulae for astronomical observations
of ``jets'' and ``blobs'' \citetext{see \citealp{SMMea97}, Appendix~A,
  and \citealp{SSO03}, Appendix~A, e.g.}.  However, what matters for
the volume transformation of a feature identified in an astronomical
image or radio map is only that the image is a supersnapshot.  What
matters for the interpretation of the supersnapshot is the volume of
fluid whose photons arrive simultaneously on the \emph{supersnapshot},
not the volume of fluid that is located within the jet volume in the
\emph{world map}.  Hence, the correct volume transformation for any
fluid volume $V\dpr$ observed by means of a supersnapshot is
\begin{equation}
V = \D\dpr V\dpr,
\label{eq:Vtrans}
\end{equation}
where $\D\dpr$ is the Doppler factor of the fluid in the observer
frame.  

If the decisive criterion was not the fact that astronomical
observations are supersnapshots, one could argue with equal
justification that the correct volume transformation formula for the
fluid in a jet section is $V\dpr = V/\Gamma$ because the jet volume is
at rest in the observer frame and hence appears contracted in the rest
frame of the fluid, or alternatively that the correct transformation
is $V\dpr = V \times \Gamma$ because the fluid is moving through the
observer frame, and therefore \emph{it} is contracted.  Both can of
course be correct, depending on whether one is \Rev{judging the jet
  volume with the help of events that are simultaneous in the fluid or
  the observer frame}. However, a supersnapshot corresponds to neither
world-map case --- the supersnapshot criterion is photons
\emph{arriving} simultaneously at the observer, which nearly always
does not correspond to photons \emph{being emitted} simultaneously in
any frame.

That eq.~(\ref{eq:Vtrans}) is correct for both the ``jet'' and ``blob''
case can be seen also by considering a section of a ``jet'' as a
collection of infinitesimal blobs that are each at rest in the fluid
frame. Alternatively, a ``jet'' can be considered as a section of a
``blob'' that is moving through a transparent gap in obscuring
material that is at rest in the observer frame --- if 90\% of a blob's
volume is covered in the observer frame, the rest-frame volume of the
visible part is 10\% of the blobs's total observer-frame volume, and
hence must also be 10\% of the blob's rest-frame volume. 

\Rev{As an alternative derivation of eq.~(\ref{eq:Vtrans}), consider that
the observer-frame volume of a ``jet'' or ``blob'' (or an
infinitesimal element of it) is given by
\begin{displaymath}
V = s \times l \times h,
\end{displaymath}
where $s$ is its extent transverse to the line of sight in the plane
of its motion, $h$ is the extent perpendicular to both the line of sight
and the direction of motion, and $l$ is along the line of sight.  The
individual factors of $V$ transform into the fluid rest frame as
follows.}

\Rev{First, since $h$ is perpendicular to the direction of motion, it is
not affected by relativity in any way, and $h\dpr = h$. Next, recall
from \S\ref{s:intro.worldsnap} above that the transverse separation
$\Delta s$ of two photon paths, i.e., light rays, is
Lorentz-invariant.  The transverse extent $s$ is defined by two such
parallel light rays and therefore it is also Lorentz-invariant, hence
$s\dpr = s$.  Finally, to determine the transformation properties of
$l$, consider the following argument. The optical depth $\tau$ along
$l$ has to be Lorentz-invariant since it encodes the fraction
$e^{-\tau}$ of photons that are absorbed by the jet material, which is
independent of the motion of any observer \citep[p.\,147]{RL79}.  By
definition, the optical depth is
\begin{displaymath}
\tau = l \, \kappa_\nu,
\end{displaymath}
where $\kappa_\nu$ the absorption coefficient of the material. The
Lorentz invariance of $\tau$ therefore implies that $l$ transforms
inversely to $\kappa_\nu$. From the Lorentz invariance of $\nu
\kappa_\nu$ \citep[again see][]{RL79}, it follows that $l$
transforms as $\nu$, i.e., $l = \D\dpr l\dpr$. Hence $ V = s \, l h =
s \, \D\dpr l\dpr \, h = \D\dpr \, s\dpr \, l\dpr \, h\dpr = \D\dpr \,
V\dpr$, again yielding eq.~(\ref{eq:Vtrans}).}

Thus, the relation between rest-frame and observer-frame volume for
supersnapshots is always given by eq.~(\ref{eq:Vtrans}), no
matter whether we are considering a ``jet'', ``blob'' or even
``shock'' feature.  As noted at the end of the preceding section, the
well-known apparent difference between the beaming formulae for a blob
and a jet \citep[$\D^{2-\alpha}$ versus $\D^3$, such as in App.\ C7
  of][]{BBR84} is in fact just a difference of \emph{integrands};
since the \emph{integration boundaries} differ depending on whether an
object is considered as blob or jet, the final answer is independent
of the assumed geometry. In other words, \textbf{jets and blobs have
  the same beaming properties if identical source volumes are
  considered}. The equivalence of jet and blob formulae will be shown
explicitly in \S\ref{s:beaming.jets.restframe} and
\ref{s:disc.Bmin.obsfixed} below.

While eq.~(\ref{eq:Vtrans}) appears straightforward to interpret, the
supersnapshot is merely a projection of the observer-frame volume onto
the plane of the sky, so that \textbf{the observer-frame volume $V$ is
  not a direct observable} (see Fig.~\ref{f:ray_sph} below). Therefore,
the volume formula can only be used for interpreting astronomical
images if an assumption is made about the geometry of the source.
However, its use in determining observables from known rest-frame
quantities is straightforward.

\begin{figure*}
 \includegraphics[width=168mm]{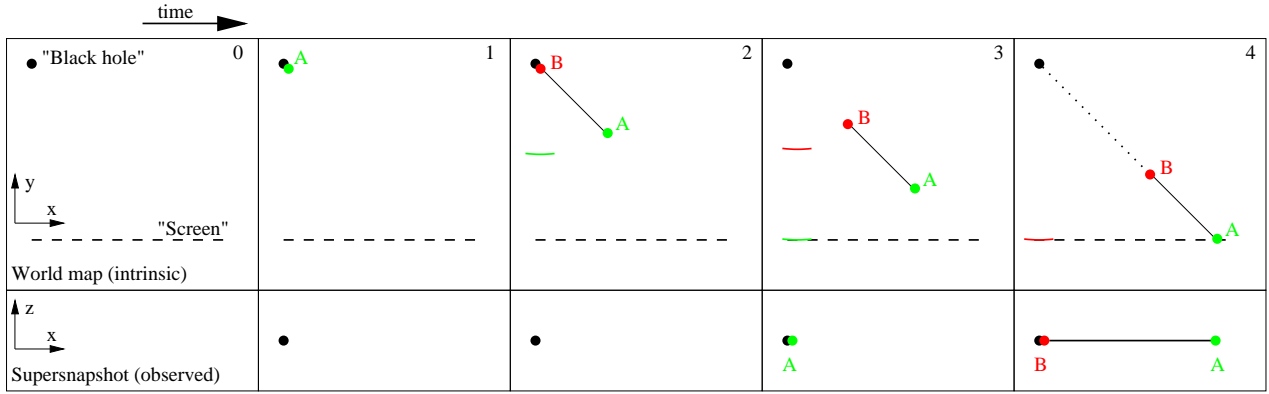}
\caption{\Rev{Illustration of retardation magnification and hiding, showing
  a sequence of events in which a relativistically moving ``blob'' is
  emitted by some source (e.g., an accretion disk around a black
  hole), and the picture recorded by a distant observer at the
  corresponding time.  The upper frames give the \emph{world map} in
  the $(x,y)$ plane with the true locations of all events; for an
  infinite speed of light, the world map corresponds to the ``top
  view'' of the events as seen by an observer at 90\degr\ to the
  blob's direction of motion. The lower frames give the
  \emph{supersnapshot}, the image projected onto the $(x,z)$ plane as
  recorded by a distant observer looking along the $+y$ axis by means
  of simultaneously arriving photons, i.e., photons that are crossing
  the dashed ``screen'' line simultaneously.  Panel \textbf{(0)} shows
  the setup, with the black dot marking the location of the source
  (``black hole'') ejecting the relativistic blob. \textbf{(1)} The
  front end ``A'' of the blob is ejected. \textbf{(2)} The rear end
  ``B'' of the blob is ejected, and at the same location as ``A'' was
  in frame (1).  ``A'' itself has travelled some distance from the
  black hole. The curved line illustrates the current location of the
  wavefront by which the observer will later imply that ``A'' has been
  ejected.  \textbf{(3)} The wavefront from the ejection of the front
  end ``A'' reaches the ``screen'' location and appears on the
  observer's picture.  The second wavefront carrying the information
  about the ejection of ``B'' is lagging behind. \textbf{(4)} The
  light from the ejection of the rear end ``B'' reaches the screen
  location.  At the same time, the front end ``A'' crosses the screen
  location.  Therefore ``B'' and ``A'' appear at the shown locations
  on the supersnapshot. The separation B--A on the supersnapshot is
  greater than it is in the world map, and the observer records a
  magnified image of the blob.  If any further material is ejected
  after ``B'' (and hence occupied the region indicated by the dotted
  line), it will not yet be visible to the observer.  Hence, the
  apparent magnification of the blob's extent implies that any further
  ejections will be unobservable until the light emitted by them has
  had time to reach the observer, thus (temporarily) being hidden from
  view.}}
\label{f:magnif}
\end{figure*}
\section{The appearance of relativistic objects}
\label{s:appearance}

This section attempts to give an intuitive pictorial representation of
how relativistic jets appear in supersnapshots.  The first part
considers jet observations as supersnapshots of infinitesimally thin
relativistic rods. This approach is appropriate for demonstrating how
the scale change in a supersnapshot, referred to below as
\emph{retardation magnification}, arises as spatial corollary to the
well-known temporal phenomenon of apparent superluminal motion
\citep{Rees66}.  The second part presents results from a ray-tracer
that demonstrate the differences between world maps and world
pictures.  Those effects are particularly important which arise from
the extent of actual jets perpendicular to the direction of motion, as
they results in extra light-travel delays between the near and far
side of the jet that are not present in the case of an idealized, thin
rod.

\subsection{Retardation magnification and hiding}

\subsubsection{Retardation magnification as corollary to
apparent superluminal motion}

Consider a relativistically moving rod, i.e., an object for which the
light-travel time \emph{across} its extent is negligible compared to
that \emph{along} its extent.  The rod is subject to Lorentz
contraction, so that its rest-frame length $\Lambda\dpr$ is related to
its length in the observer frame by
\begin{equation}
\Lambda = \Lambda\dpr / \Gamma,
\label{eq:Ltrans}
\end{equation}
where $\Lambda$ is inferred from a world-map analysis.  For the
interpretation of observations, where the observer is sufficiently
distant from the moving rod to be able to take a \emph{supersnapshot},
we want to relate the apparent size of the object on the supersnapshot
to its actual length in the world map. To clarify terminology, the
term ``projected size'' (symbol $\Lambda\proj$) will be used to mean
the size that corresponds to the projected extent as measured on the
supersnapshot, while the term ``apparent size'' ($\Lambda\app$) is the
size that is inferred from the projected size by deprojecting with the
line-of-sight angle $\theta$, i.e., $\Lambda\proj = \Lambda\app \sin
\theta$.  From the constancy and finite value of the speed of light,
\begin{eqnarray}
\label{eq:Ltrans_mag} 
\Lambda\app &=& \frac{\Lambda}{1-\beta \mu} \\
&= & \magfac \Lambda, \label{eq:Ltrans_magfac}
\end{eqnarray}
where I have defined a magnification
\begin{equation}
\magfac = (1-\beta\mu)^{-1}.
\label{eq:magfac}
\end{equation}
Substituting eq.~(\ref{eq:Ltrans}) recovers the well-known
relation
\begin{equation}
\Lambda\app = \D \Lambda\dpr.
\label{eq:Ltrans_D}
\end{equation}
\citep[e.g.,][eqn.\ {[12]}]{Ghis00}.  Equivalently,
\begin{eqnarray}
\Lambda\proj & = & \D \Lambda\dpr \sin \theta \label{eq:Ltrans_Dproj} \\
 & = & \frac{\Lambda \sin \theta}{1-\beta \mu} \label{eq:Ltrans_Dproj_obs} \\
 & = & \magfac\,\Lambda \sin \theta
\label{eq:Ltrans_appobs}
\end{eqnarray}
Not by coincidence, the projected velocity $v\proj$ (usually called
``apparent transverse velocity'') in apparent superluminal motion is
related to the actual velocity $v$ by exactly the same magnification
factor that relates the projected length to the actual length:
\begin{eqnarray}
v\proj & =&  \frac{v \sin \theta}{1-\beta\mu} \nonumber \\
&=& \magfac\, v\sin\theta. \nonumber
\end{eqnarray}
From eqs.~(\ref{eq:Ltrans_mag}), (\ref{eq:Ltrans_Dproj}) and
(\ref{eq:Ltrans_Dproj_obs}), we can read off all the relevant
implications for the interpretation of supersnapshots:
\begin{enumerate}
\item For an approaching rod with $\mu>0$, $1-\beta\mu < 1$ and
  therefore $\magfac > 1$ for any value of $\beta$.  Thus, the
  \emph{apparent}, deprojected length $\Lambda\app$ of a relativistic
  approaching object is always greater than its observer-frame length
  $\Lambda$ as inferred from a world map.  In other words, \textbf{in
    a supersnapshot, any approaching relativistic object appears
    magnified compared to its actual observer-frame size}. 
  Only for $\theta=90\degr$, $\mu=1$ and
  $\Lambda\proj=\Lambda\dpr/\gamma=\Lambda$, and the Lorentz
  contraction of a thin rod becomes observable.
\item The projected length $\Lambda\proj$ is always less than or equal
  to the rest-frame length $\Lambda\dpr$. The limiting case
  $\Lambda\proj=\Lambda\dpr$ occurs for $\mu=\beta$, which implies
  $\D=\gamma$ and $\sin\theta = 1/\Gamma$. In this case, the
  ``projected'' appearance of the rod in a supersnapshot is identical
  to the view of an observer looking at the rod from 90\degr\ in its
  own rest frame, without any Lorentz contraction.  This fact was
  already derived at the end of \S\ref{s:intro.worldsnap}, there based
  on eq.~(\ref{eq:sintrans}). 
\end{enumerate}
The second point implies that it is possible to constrain the
orientation and speed of a moving relativistic object if its
rest-frame size is known. \Rev{In the case of jets, this may be
  possible if jet features are known to have a certain ratio of length
  to width, since the width is not affected by the beaming.}

Figure~\ref{f:magnif} gives a more intuitive illustration of why
distant relativistic objects appear magnified in a supersnapshot
compared to their true observer-frame extent.  As in apparent
superluminal motion, the cause of this effect is the time delay
between light signals reaching the observer from the end of the object
that is closest to the observer and those from the end of the object
furthest from the observer.  Therefore, I call this effect
\emph{retardation magnification}.  In analogy to the first point
above, any relativistic object's apparent velocity is \emph{always
  magnified} compared to its true velocity, without necessarily
appearing to be superluminal.

\subsubsection{Retardation hiding in astrophysical jets}

For astrophysical jets, however, there is a catch: jets are produced
by accretion disks around a compact object \citetext{``core''; for
  active galactic nuclei, the compact object is a black hole, and
  similar jets are launched from accretion disks around black holes
  [\citealp{Liebovitch74}, e.g.], neutron stars and white dwarfs in
  X-ray binaries and novae [see \citealp*{FBG04}, e.g.]}, and they terminate
in a ``hot spot'' (this can be a shock terminating an FR~II jet, or a
flaring point at which an FR~I jet decelerates substantially).  Both
are moving through the observer frame much more slowly than the jet
material itself.  Hence, the apparent size of any jet feature cannot
be larger than the separation between the core and the hot spot.

Thus, if jet features appear magnified, it means that not all of the
features that are actually \emph{present} in the observer frame can be
\emph{visible simultaneously} to the observer, as the total apparent
length would then need to be larger than the actual separation between
core and hot spot. Hence, the magnification implies that parts of the
jet are hidden from the observer's view.  This retardation
\emph{hiding} occurs because light signals from the further end of the
jet have not yet had time to reach the distant observer, even though
they have already emerged from the core.  As the object appears
magnified by a factor $\magfac$, it follows that only a fraction
$1/\magfac = 1-\beta\mu$ of the object is visible; since time delays
are greater for the parts of the object that are further from the
observer, the visible part of the jet is that closest to the observer,
and the hidden part is that furthest from the observer.
\begin{figure}
\includegraphics[width=84mm]{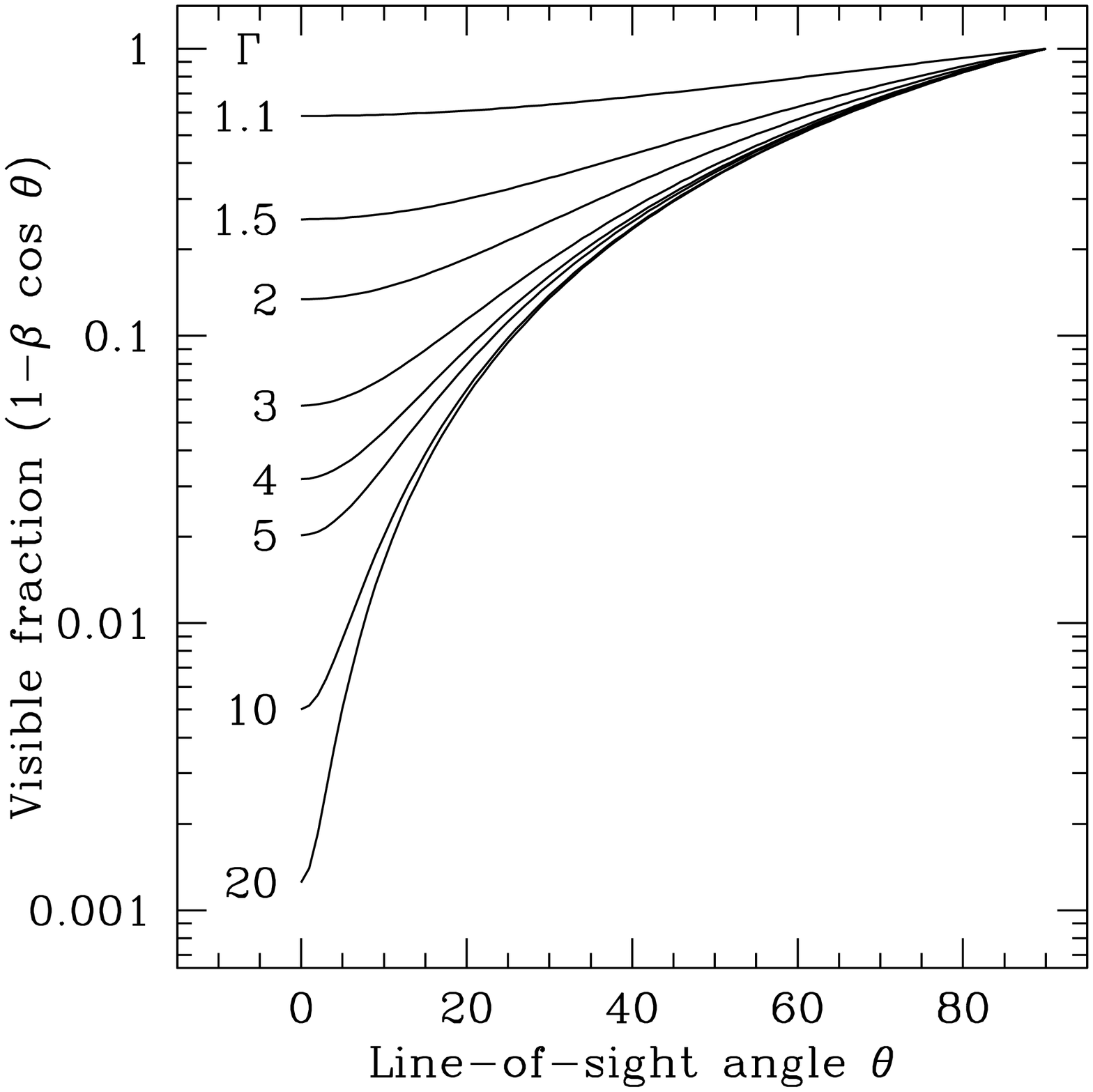}
\caption{Visible fraction of a jet, or other relativistically moving
  body emerging from a stationary one and disappearing into another
  stationary one, on a \emph{supersnapshot}, as function of the angle
  $\theta$ between the observer's line of sight and the jet's
  direction of motion, for different Lorentz factors $\Gamma$ as
  given.  The visible fraction is the inverse of the magnification
  factor $\magfac = (1-\beta\mu)^{-1}$ defined in the
  text. Relativistic flux beaming affects the detectability of jets;
  for $\Gamma \ll 1$, the visible fraction of a jet at the critical
  beaming angle $\theta \approx 1/\Gamma$ is given by $5/(8\Gamma^2)$;
  as the jet becomes significantly de-beamed for larger line-of sight
  angles, this is the maximum visible fraction of the jet material in
  a relativistic jet.}
\label{f:visfrac}
\end{figure}
\begin{figure*}
\includegraphics[width=50mm]{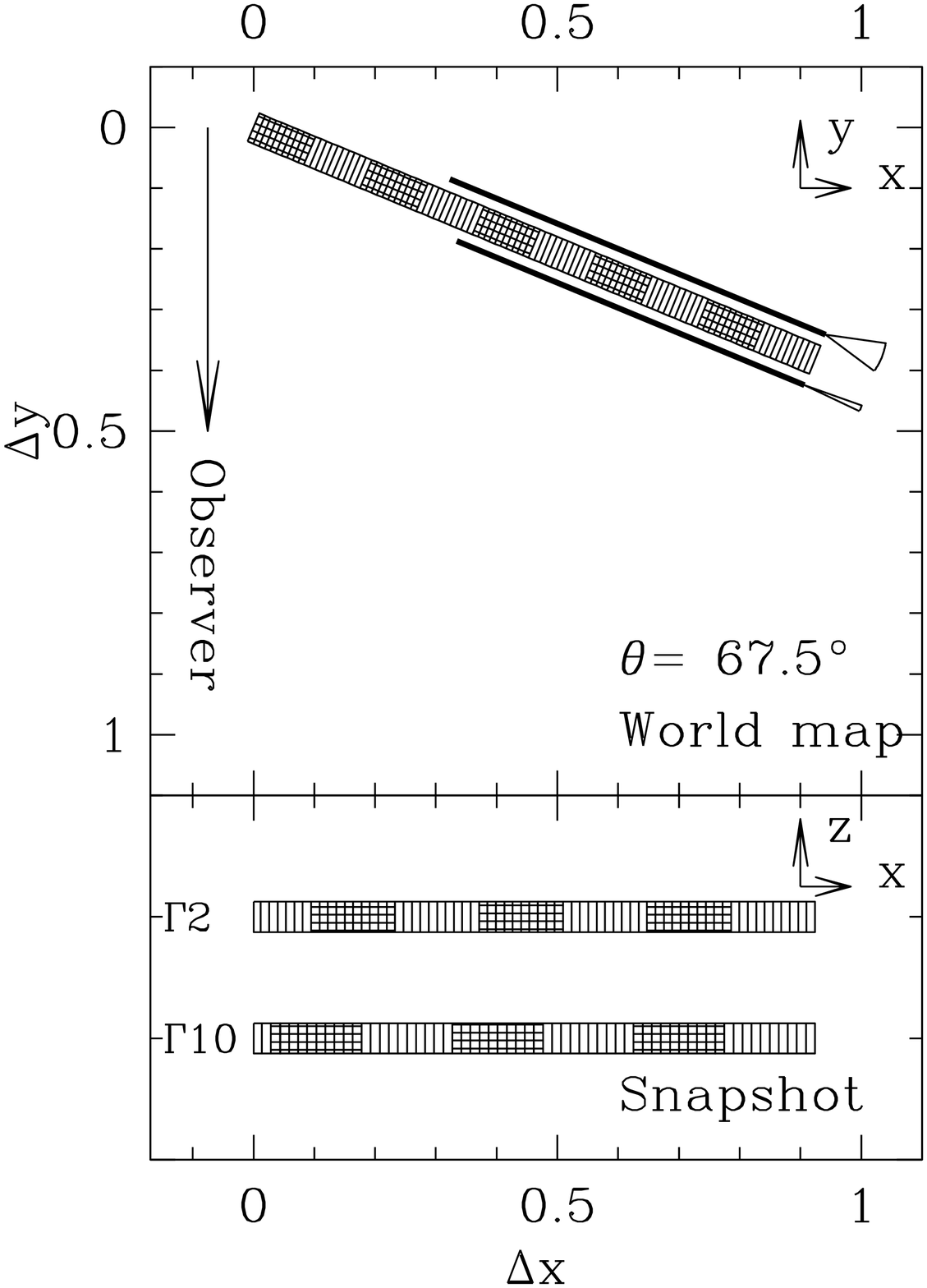}\hfill
\includegraphics[width=50mm]{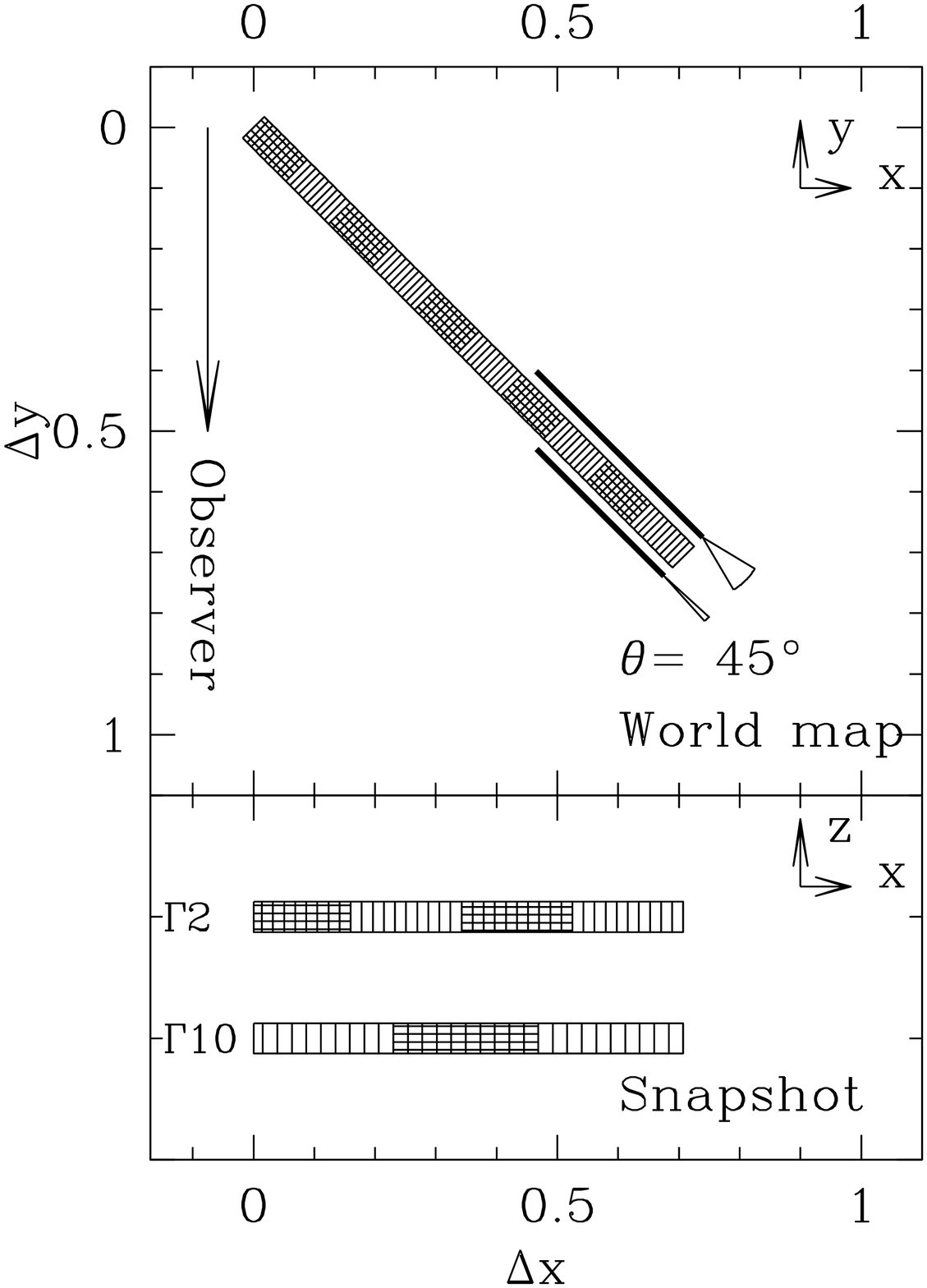}\hfill
\includegraphics[width=50mm]{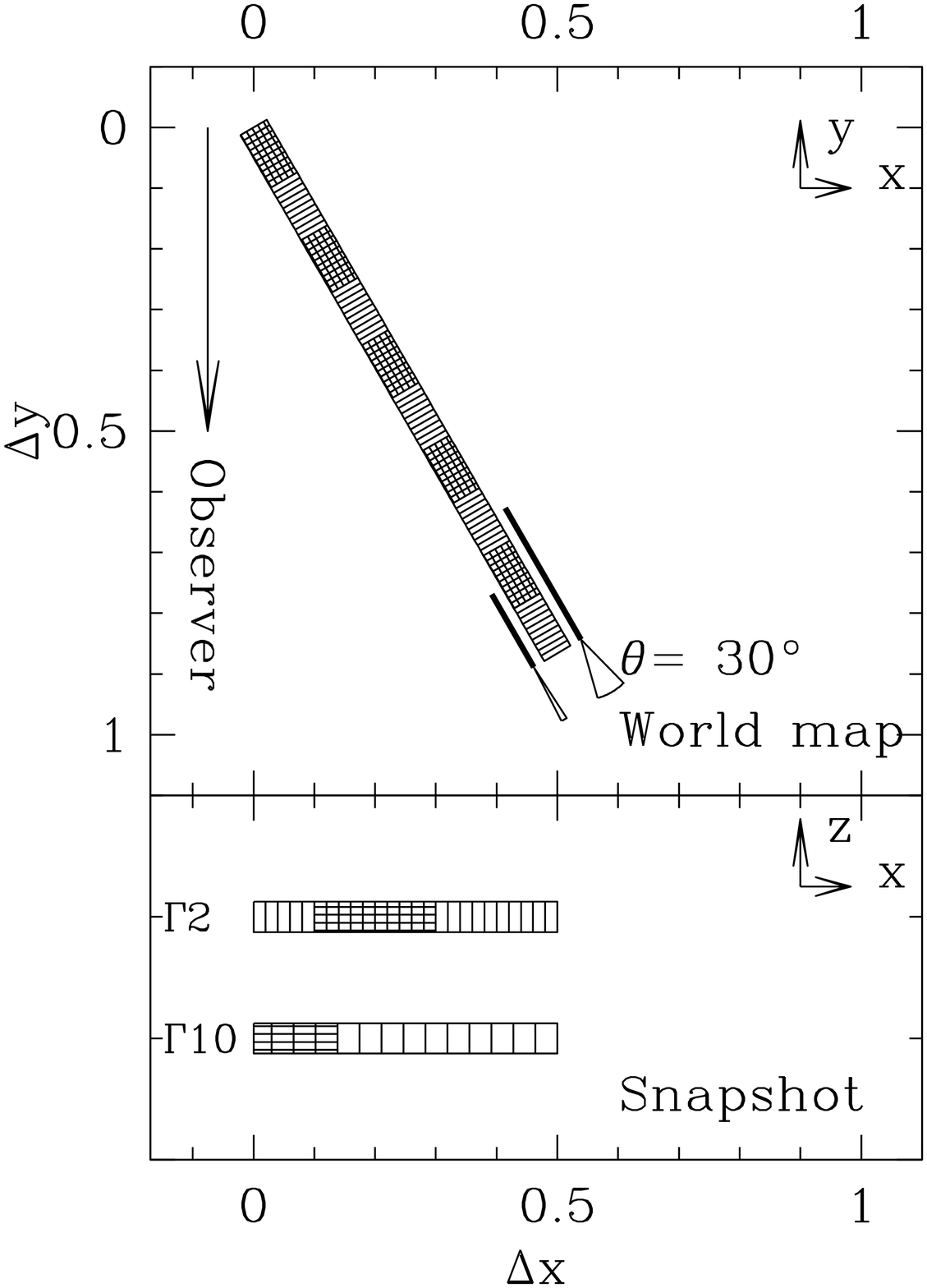}
\includegraphics[width=50mm]{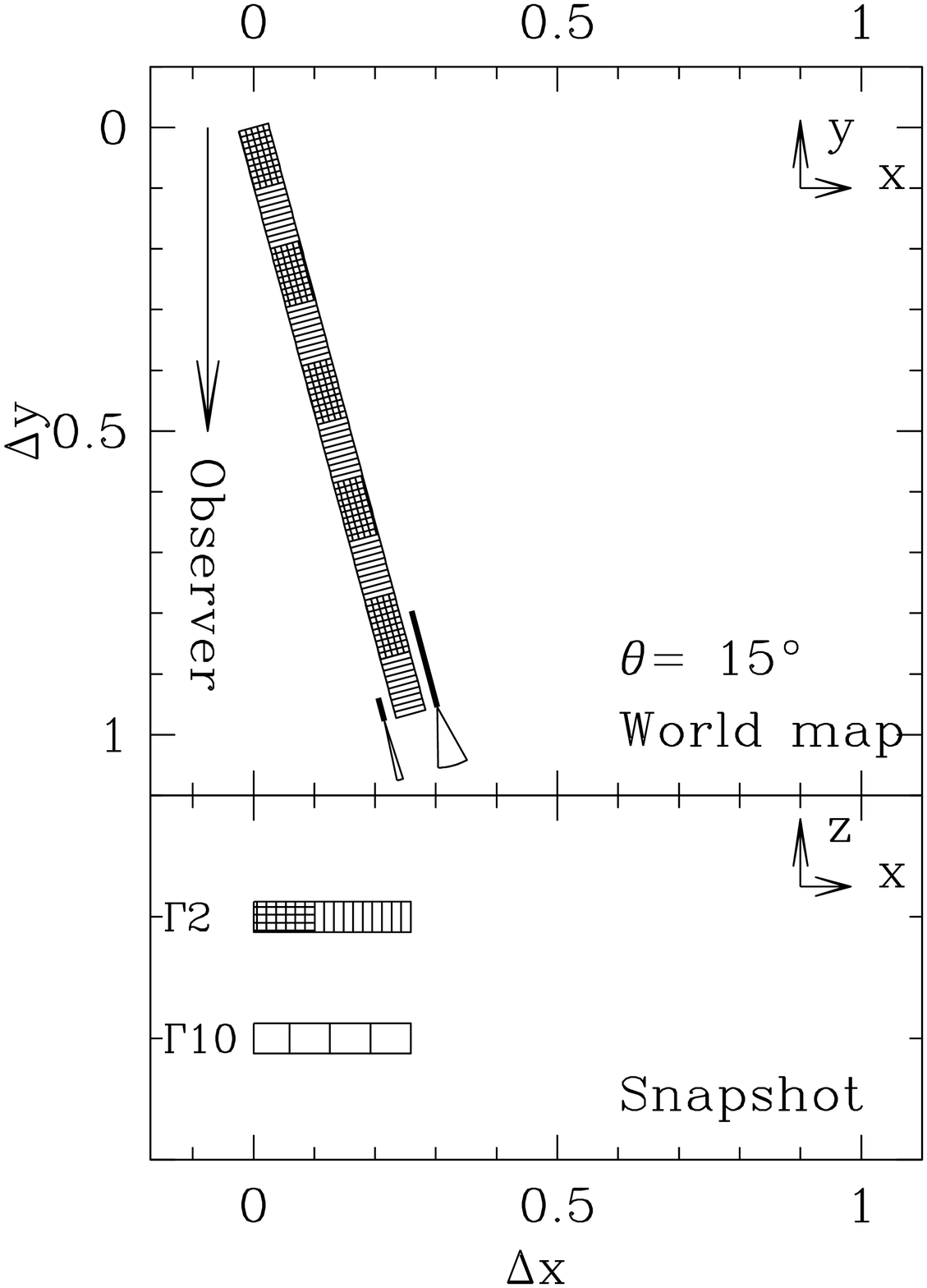}\hfill
\includegraphics[width=50mm]{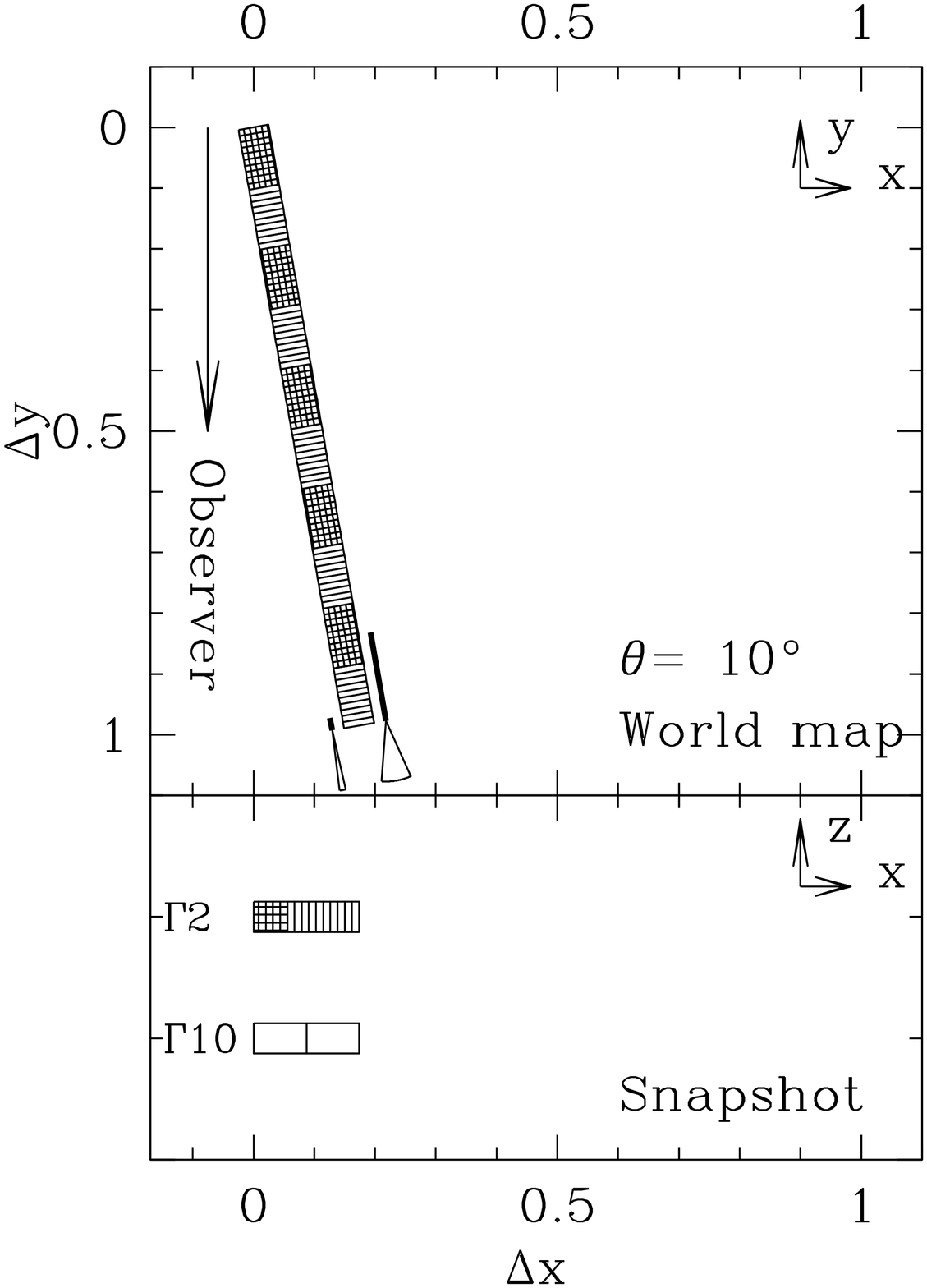}\hfill
\includegraphics[width=50mm]{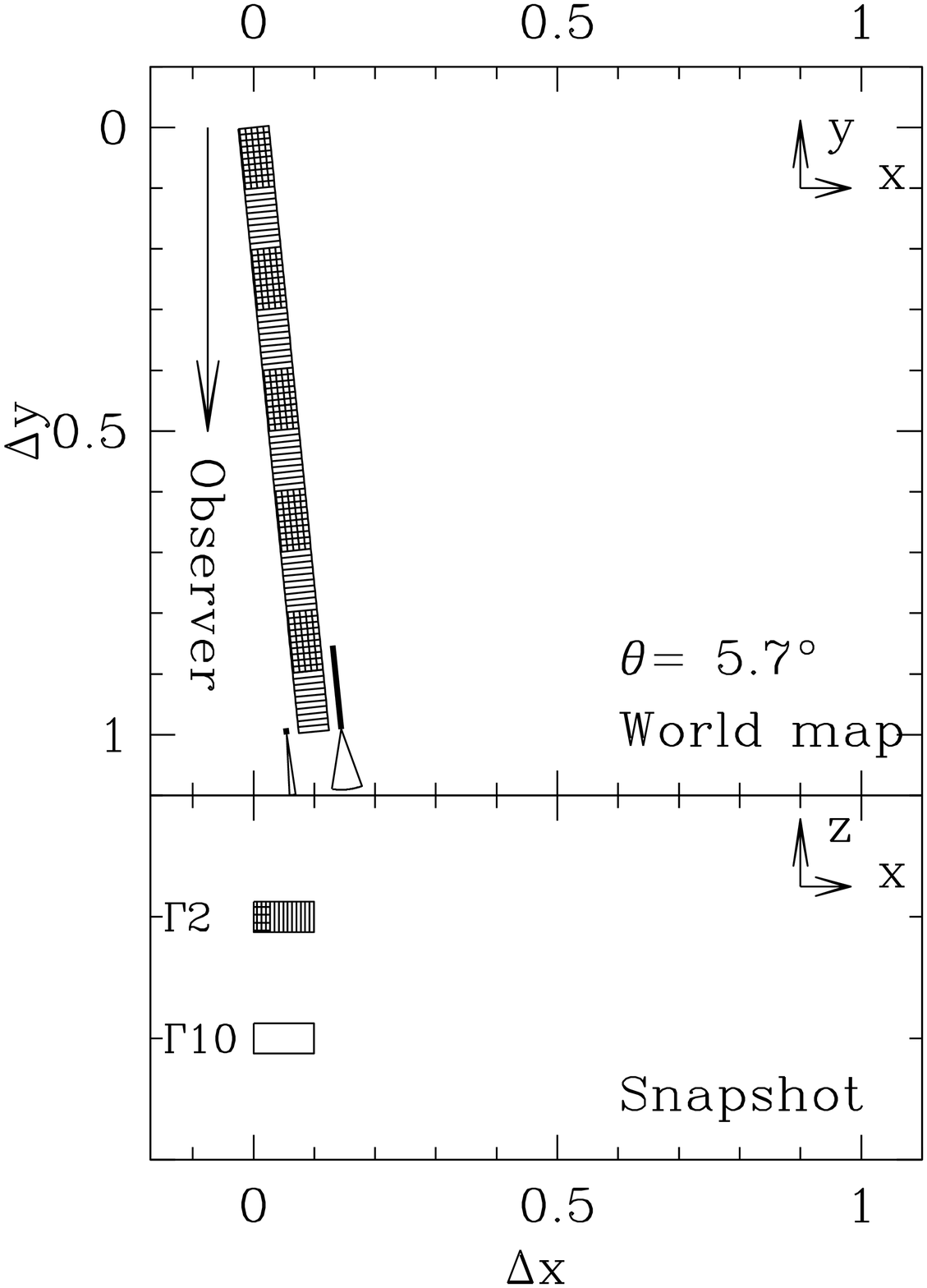}
\caption{Visual appearance of a relativistic rod with features that
  are fixed in the object's rest frame (i.e., 1-dimensional blobs),
  for different line-of-sight angles $\theta$ and Lorentz factors. As
  in Fig.~\protect\ref{f:magnif}, the upper panel in each plot shows a
  ``top view world map'', while the lower panel shows the appearance
  of the jet in a supersnapshot taken by an observer at $y=-\infty$.  The
  solid bars above and below the jet in the upper panels indicate the
  fraction $1-\beta\mu$ of the jet that is visible to the pole-on
  observer, above the jet for $\Gamma=2$ (longer bar) and below the
  jet for $\Gamma=10$.  Those photons arriving at a projected position
  just next to the core were emitted by jet material adjacent to the
  inner end of the bar at the time when it was just next to the core.
  All parts of the jet that are closer to the core are not yet visible
  because the photons from those parts of the jet have not yet had
  time to reach the observer.  The sectors at the ends of the solid
  bars indicate the relativistic beaming cone of half-opening angle
  $\Gamma/2$; for the angles and Lorentz factors shown here, only the
  jets with $\Gamma=2$ at angles $\theta\leq15\deg$ have their fluxes
  enhanced by beaming, while the remainder have their fluxes
  significantly suppressed by beaming.}
\label{f:angles}
\end{figure*}
Figure~\ref{f:visfrac} shows the visible fraction as function of
Lorentz factor and line-of-sight angle. Figure \ref{f:angles} illustrates
which parts of a relativistic rod are visible to a distant observer.

The relativity-textbook analogue would be a relativistic train
emerging from one tunnel and disappearing into a second one.  If the
distance between the end of one tunnel and the beginning of the second
one is sufficiently short, a distant observer looking at the train
from a small angle to the train's direction of motion will observe
fewer railway carriages between the two tunnels than actually fit
between them as judged by observers creating a world map.

The magnification obviously applies to blobs, as their outlines are at
rest in the \emph{fluid} frame. However, it also applies to individual
fluid elements that make up a ``jet'' feature whose outline is at rest
in the \emph{observer} frame.  Hence, a given section of an
approaching jet actually \emph{contains} more fluid elements than are
\emph{visible} simultaneously to an observer.  This is another way of
deriving that eq.~(\ref{eq:Vtrans}) is always the correct volume
transformation for supersnapshots, even in the jet case.

The discussion so far has dealt exclusively with 1-dimensional rods,
with negligible light travel time in the direction transverse to the
direction of motion compared to the light travel time along the
direction of motion. This situation does not apply to features of real
astrophysical jets, which have comparable extent along and across the
direction of motion.  The next section illustrates the effect of
light-travel delays across a relativistically moving object.

\subsection{Ray-tracing simulations of supersnapshots}
\label{s:illust.rays}

To give a visual illustration of the difference between a world map
and a supersnapshot, this section contains images of the same emitting
regions, once in the world map view, identical to what would be seen
if the speed of light was infinite, and once in a supersnapshot view,
appropriate for astronomical observations.  As just noted, the
discussion in the preceding section applies directly to ``blob''
features, whose outline is at rest in the fluid frame, but also to
individual fluid elements making up a ``jet'' feature, whose outline
is at rest in the observer frame.  The main difference between a blob
and a jet for the ray-tracing images is that the \emph{outline} of
blobs experiences the magnification effects. By contrast, the outline
of stationary jet features does not get magnified and behaves
according to our intuition, which is formed by observing bodies moving
at velocities much less than the speed of light.  Hence, this section
concentrates on the aspect of how the observed (projected) and true
geometry of objects in supersnapshots are related, including the
question how to infer the volume of a relativistic object from a
supersnapshot.  

The discussion here and in the entire the paper is restricted to
optically thin objects.  The effects of relativistic beaming on flux
measurements and quantities derived from them will be examined in the
following section.

\begin{figure*}
\includegraphics[width=.65\hsize]{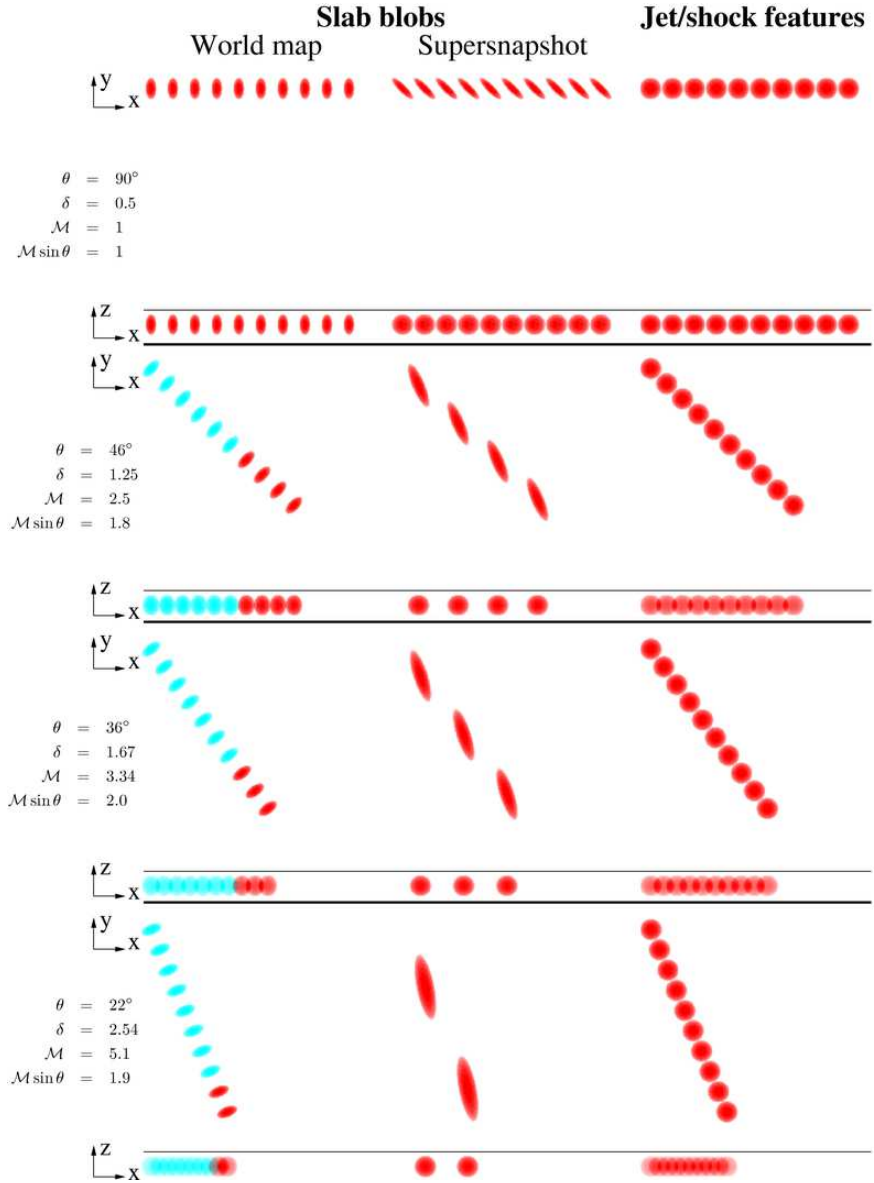}
\caption{\label{f:ray_sph}World maps and supersnapshots of spherical,
  optically thin blobs and stationary jet features, with the fluid
  moving with Lorentz factor $\Gamma=2$ at the line-of-sight angles as
  shown.  The two leftmost columns show blobs which are spherical in
  their own rest frame, the right-hand column shows spherical jet
  features.  The observer is located at $y=-\infty$, and blobs that
  are visible to the observer are shown in red, while blobs that are
  hidden to the observer are shown in cyan.  As in previous figures,
  there is a panel each for a ``top view'' and the projected
  appearance in the plane of the sky; in the case of the
  ``supersnapshot'', both projections include light-travel delays
  towards the observer, i.e., photons originating at larger $y$ values
  left the source at progressively earlier times.  For the
  ``supersnapshot'' and the ``jet'' column, the projection onto the
  plane of the sky is identical to the supersnapshot obtained by a
  distant observer (for the jet case, it is assumed that the jet is
  much older than the light travel time difference from the furthest
  to the nearest feature); for the ``world map'' column, the
  projection on the plane of the sky is what \emph{would} be seen if
  the speed of light was infinite.  The world map shows where the
  emitting material actually \emph{is}, while the supersnapshot shows
  what is \emph{observed}.  The salient points about this figure are:
  \textbf{(1)} spherical blobs always appear as spherical blobs, but
  their true observer-frame shape is ellipsoidal; moreover, spherical
  blobs are magnified \emph{along} the line of sight, retaining the
  orientation-dependent volume $V\app=\D V\dpr$, but the projection
  onto the plane of the sky makes this effect unobservable as far as
  the shape of the object is concerned; \textbf{(2)} while the
  \emph{shape} of the spherical blobs is unaffected by changes in
  $\D$, their \emph{spacing} is affected; the projected length of the
  entire sequence of blobs is the same in the supersnapshot and the
  world map, but taken up by fewer blobs in the snapshot than there
  actually are; therefore, some of the blobs are hidden (the hidden
  blobs are shown in cyan).  Thus, \textbf{(3)} the projected
  appearance of jets and blobs behaves differently as the
  line-of-sight angle is reduced: jet features move closer together
  and begin to overlap as the line of sight passes through multiple
  features, while blobs appear to move further apart.}
\end{figure*}
\begin{figure*}
\includegraphics[width=.65\hsize]{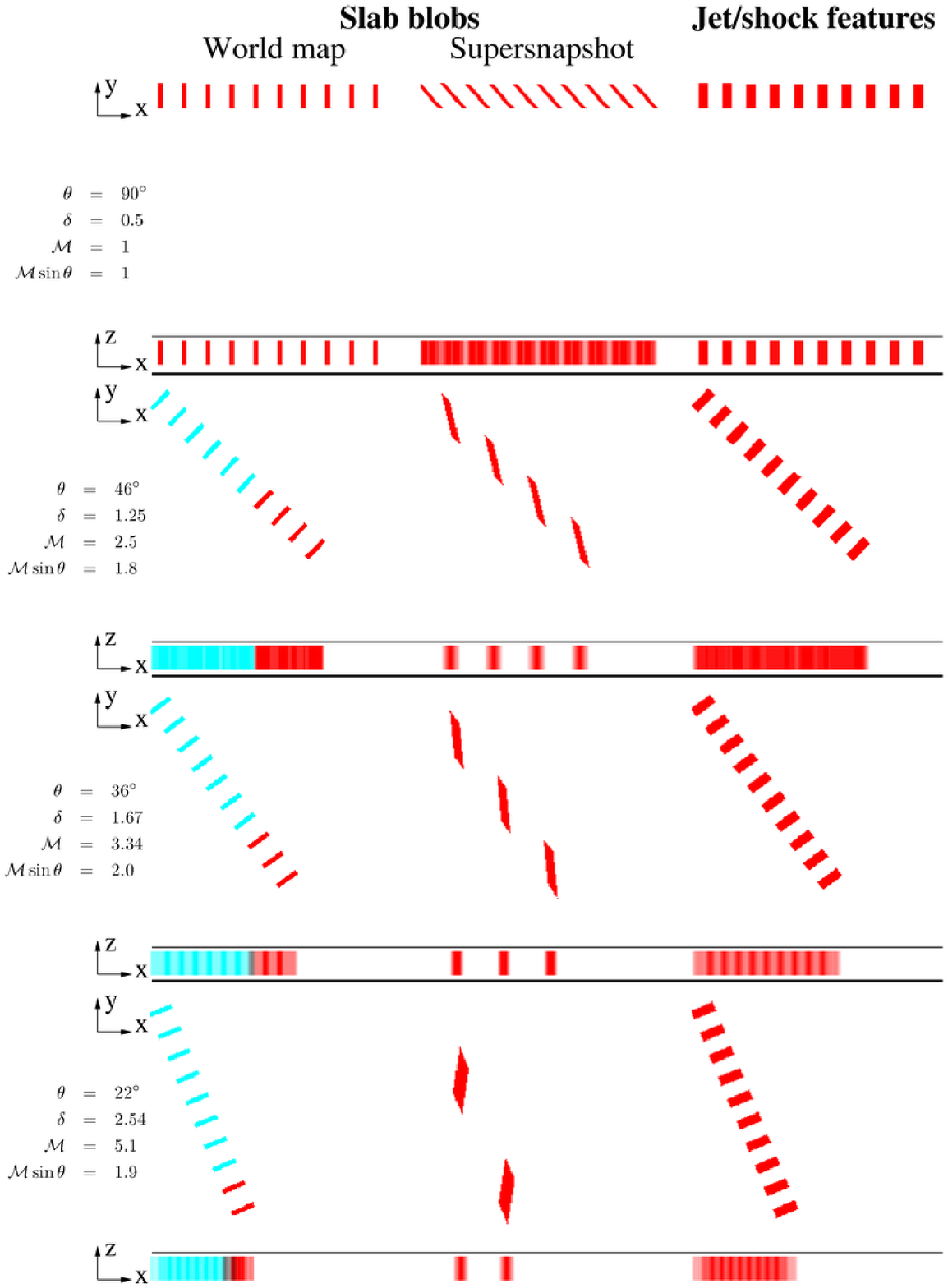}
\caption{\label{f:ray_slab}As Fig.~\protect\ref{f:ray_sph}, but with
  slab features of rectangular cross-section. In the slab geometry,
  the change of the blob-frame line-of-sight angle is seen clearly,
  because the projected appearance now varies with that angle: at
  observer-frame line-of-sight angles greater than the critical angle
  given by $\sin\theta = 1/\Gamma$ (corresponding to the ``beaming
  cone'' half-opening angle), i.e., $\theta = 30\degr$ for $\Gamma=2$,
  even approaching blobs are seen from behind in their own rest-frame,
  and only at line of sight angles $\theta < 30\degr$ are the jet
  features seen from the front. In fact, for a wide range of angles,
  the features are seen nearly side-on in their own rest frame
  \protect\citep[compare][]{BB96}. As in the spherical-blob case,
  fewer and fewer blobs are visible for smaller values of $\theta$,
  and the projected spacing of the knots remains constant over the
  same range of angles where the knots are seen roughly edge-on. }
\end{figure*}
Figures \ref{f:ray_sph} and \ref{f:ray_slab} shows some simple
ray-tracing pictures illustrating the difference between world map and
supersnapshot for blobs, and the different projection properties for
relativistically moving and stationary features.  The following points
about observations blobs or jets can be inferred from
Figs.~\ref{f:ray_sph} and \ref{f:ray_slab}:
\begin{enumerate} 
\item As expected from \citet{Pen59}, blobs that are spherical in
  their rest frame are always observed as spherical blobs, even though
  they are in fact lens-shaped in the observer frame (due to the
  Lorentz contraction along the direction of motion). They are still
  magnified in the supersnapshot, but the magnification is now
  \emph{along} the line of sight; this is why their observer-frame
  volume still scales as $V\app=\D\dpr V\dpr$, even though their
  projected appearance is identical to their rest-frame appearance.
\item The observed \emph{shape} of spherical blobs is not affected by
  changes in the line-of-sight angle, and hence the Doppler factor;
  however, their \emph{spacing} changes by a factor $\D\dpr$ with
  respect to their rest-frame spacing, by $\magfac$ with respect to
  their true observer-frame spacing, resulting in an observed
  projected spacing of $\magfac\sin\theta$ times their true
  observer-frame spacing.  Thus, relativistic blobs (and relativistic
  objects in general) behave very unintuitive under changes of the
  observation angle -- the closer the line-of-sight angle is to the
  critical angle given by $\sin\theta = 1/\Gamma$ to the direction of
  motion, the further they appear to be apart.
\item By contrast, stationary spherical jet features behave as
  expected by our everyday intuition, with smaller projected spacings
  between individual features for smaller line-of-sight angles.
\item The geometry of individual blobs (spherical vs.\ slab) does not
  affect \emph{how many} of the blobs are seen, or their total beamed
  luminosity, but it does affect how the \emph{projected appearance}
  of individual blobs varies with line-of-sight angle.
\end{enumerate}
The figures do not consider moving shocks, features which are at rest
neither in the frame of the observer nor in that of the
fluid. However, their projected shape as function of Doppler factor is
identical to the projected shape of blobs, i.e., it is governed by the
shock's Doppler factor $\D\pr$. The fluid's emissivity in the observer
frame, on the other hand, is governed by the fluid's Doppler factor
$\D\dpr$.

\section{Relating observed and rest-frame quantities in jet observations}
\label{s:beaming}

Beaming formulae have been given in numerous places in the literature.
The usual approach is one that considers a \emph{fixed source volume}
and then determines the changes in the received flux density, surface
brightness etc.\ resulting from changes in the Doppler factor, i.e.,
the line-of-sight angle and Lorentz factor. However, in astronomical
observations, it is not the \emph{source} volume that is fixed, but
the source's \emph{projected appearance} in the supersnapshot.  The
fact that the source volume differs when deprojecting the observed
appearance with different line-of-sight angles is not considered often
in the beaming literature.  It is therefore worthwhile to restate the
beaming prescriptions for both directions, inferring changing
observables for fixed source properties, and changing source
properties from fixed observables.  As above, the equations assume
optically thin, isotropic emission with a power-law emissivity
$j_{\nu} \propto \nu^{\alpha}$ that is constant within the emitting
region.

\subsection{Beaming and de-beaming blobs}
\label{s:beaming.blobs}

In the case of a blob, the pattern frame is identical with the fluid
frame, so that $\D\pr = 1$ and $\D\dpr$ is the relevant Doppler
factor.  To get observed quantities in terms of source parameters, we
need to specify a geometry for the emission region.  I consider simple
spherical emission regions as well as ``slab''-shaped regions of
either square or circular cross-section.

\subsubsection{Flux beaming of blobs}
\label{s:beaming.blobs.sphflux}

The beamed total flux density of a blob is given by
\begin{equation}
S_\nu(\nu) = \dL^{-2} \D\dpr^{3-\alpha} j\dpr_{\nu\dpr}(\nu)\; V\dpr,
\label{eq:blobfluxdens}
\end{equation}
where emissivity and source volume are held fixed in the fluid rest
frame. The beamed bolometric luminosity is obtained by integrating
$S_\nu$ over the appropriate range of \emph{rest-frame} frequencies,
leading (after substitution using eqs.~\ref{eq:Ldef} and
\ref{eq:nutrans}) to
\begin{equation}
L = \D\dpr^4 L\dpr.
\label{eq:bloblum}
\end{equation}
In general, the beaming properties of bolometric quantities are
obtained from the frequency differentials by adding $1+\alpha$ to the
exponent of the Doppler factor, and the exponent to the Doppler factor
of beamed bolometric quantities does not contain any spectral-shape
parameter such as $\alpha$.  

\subsubsection{Surface brightness beaming of blobs}
\label{s:beaming.blobs.SB}

Surface brightness beaming is important because the detectability of
jet features in optical and radio observations is determined by their
peak surface brightness, not by their total flux.  The observed
surface brightness is given by the rest-frame volume emissivity
integrated along the observer's line of sight after rotation into the
rest frame.

\paragraph{Spherical blobs}

For a spherical blob, the beamed surface brightness is
$\D\dpr^{3-\alpha}$ times the surface brightness that would be
perceived by an observer at rest with respect to the blob, but at the
same distance.  Since the surface brightness of a spherical blob
varies according to the different length of the line of sight as
function of sky coordinates, I do not give an explicit formula here.

\paragraph{Slab blobs}

For a slab blob with square cross-section of side length $w$, the
observed surface brightness far enough from the edges is independent
of sky coordinates and given by
\begin{equation}
I_\nu = \frac{\D\dpr^{2-\alpha}}{\sin \theta} j\dpr_{\nu\dpr}(\nu)\;w.
\label{eq:slabblobsurfbright}
\end{equation}
Ignoring edge effects is appropriate for slab blobs with length:width
ratios greater than about 2. The same expression applies to the
surface brightness along the projected axis of a cylindrical blob with
observed diameter $w$, and the scaling of observed surface brightness
with the beaming parameters $\Gamma,\theta$ applies to the entire
cylindrical blob.

\subsubsection{Inferring the rest-frame properties of blobs}
\label{s:beaming.blobs.restframe}

Equations (\ref{eq:blobfluxdens}) and (\ref{eq:bloblum}) by themselves
give the dependence of received flux and inferred luminosity on
observation angle and Lorentz factor.  In order to infer rest-frame
quantities such as $j\dpr$ from observations, it is necessary to infer
the rest-frame volume from the source's projected appearance.  The
prescription for this depends on source geometry.

\paragraph{Spherical blobs}

As discussed above, a spherical blob will appear as the same spherical
blob to any observer, including the rest observer. Hence, the
rest-frame volume is simply $V\dpr = 4/3 \pi R^3$, where $R$ is the
observed radius of the blob.  This apparently contradicts
eq.~(\ref{eq:Vtrans}), but there is in fact no contradiction, as can
be seen from Fig.~\ref{f:ray_sph}: the $(x,y)$ view of the
``supersnapshot'' column shows that for $\D>1$, the (unobservable)
``observer-frame volume'' of a spherical blob has a larger extent
\emph{along} the line of sight than can be inferred from its projected
appearance, and a smaller one for $\D<1$. Thus, the rest-frame
emissivity of the material is given by
\begin{equation}
j\dpr_{\nu\dpr}(\nu) = S_\nu(\nu) \,\dL^{2}
\frac{\D\dpr^{-3+\alpha}}{4/3\pi R^3}.
\label{sphblob_jtrans}
\end{equation}

\paragraph{Slab blobs}

Next, I consider a slab-shaped blob with square or circular
cross-section $A$, i.e., a blob whose intrinsic size perpendicular to
the direction of motion can be inferred directly from its transverse
angular size in the plane of the sky.  The rest-frame length of the
slab can be inferred from the projected length $\Lambda\proj$ and the
fact that our line of sight crosses the blob in its rest frame at an
angle given by eq.~(\ref{eq:sintrans}), i.e., $\Lambda\dpr =
\Lambda\proj/(\D \sin\theta)$.  Thus, $V\dpr = A\,\Lambda\proj/(\D
\sin\theta)$ and
\begin{equation}
j\dpr_{\nu\dpr}(\nu) = S_\nu(\nu) \,\dL^{2}
\frac{\D\dpr^{-2+\alpha}\,\sin\theta}{A \Lambda\proj},
\label{slabblob_jtrans}
\end{equation}
again ignoring edge effects.

Comparing to the relation between observed flux and rest-frame volume
emissivity, eq.~(\ref{eq:blobfluxdens}), it is appears to be a
contradiction that eq.~(\ref{slabblob_jtrans}) has a different
scaling with Lorentz factor and observation angle. The difference
arises because eq.~(\ref{eq:blobfluxdens}) applies to
observations of the same \emph{source} from different directions
($V\dpr$ is kept fixed), while eq.~(\ref{slabblob_jtrans})
describes the situation where the \emph{observables} are kept fixed,
so that different line-of-sight angles correspond to different
inferred values for $V\dpr$. The decisive aspect that is often
neglected is the length of the sight line across the blob changes due
to the relativistic angle aberration (eq.~\ref{eq:sintrans}), leading
to the $\sin\theta$ term in the final expression.  Because of this
line-of-sight deprojection, the ``de-beaming'' formula cannot be
expressed as function of the Doppler factor $\D\dpr$ only.

\subsection{Beaming and de-beaming jets}
\label{s:beaming.jets}

\subsubsection{Beamed flux density of a jet}
\label{s:beaming.jets.flux}

The beamed flux density of a jet feature is given by
\begin{equation}
S_\nu(\nu) = \dL^{-2} \D\dpr^{2-\alpha} j\dpr_{\nu\dpr}(\nu)\;V.
\label{eq:jetfluxdens}
\end{equation}
This arises simply from integrating the fluid-frame emissivity
transformed into the observer frame over the observer-frame volume.
Though comparison with the blob case, eq.~(\ref{eq:blobfluxdens}),
seems to reveal the usual difference in the exponent of the Doppler
factor between ``jet'' and ``blob'' case, the expression is multiplied
by \emph{different} volumes, fluid-frame $V\dpr$ for the blob case,
observer-frame $V$ for the jet case, that are at rest in
\emph{different} frames. Therefore, merely comparing the exponent of
the Doppler factor does not contain the full information about beaming
properties of blobs versus jets.

\subsubsection{Beamed surface brightness of a slab jet}
\label{s:beaming.jets.SB}

If the emitting volume is a slab jet with transverse width $w$, the
volume of a section with projected length $\Lp$ is just $V=w^2 \Lp /
\sin\theta$.  Hence, the observed surface brightness of such a slab
jet is
\begin{equation}
I_\nu = \frac{\D\dpr^{2-\alpha}}{\sin \theta} j\dpr_{\nu\dpr}(\nu)\; w,
\label{eq:slabjetsurfbright}
\end{equation}
This is identical to the beamed surface brightness of a slab
\emph{blob} (eq.~\ref{eq:slabblobsurfbright}), as it must, because a
continuous jet can be considered as a section of a long blob.  

\subsubsection{Inferring rest-frame properties of jets}
\label{s:beaming.jets.restframe}

Asking about the ``rest-frame'' volume of a jet feature is not
directly meaningful, since the outline of the jet feature is at rest
in the observer's frame, while the emitting fluid is not, so there
isn't a single frame in which both the fluid and its outline are at
rest.  However, by substituting eq.~(\ref{eq:Vtrans}), we can of
course express eq.~(\ref{eq:jetfluxdens}) in terms of the volume
$V\dpr=V/\D\dpr$ of jet fluid that contributes photons to the
observer's supersnapshot, obtaining the same expression as for the
``blob'' case, eq.~(\ref{eq:blobfluxdens}), and hence also the
same expression for the relation between observer-frame and
fluid-frame bolometric luminosity, eq.~(\ref{eq:bloblum}).

If the supersnapshot nature of astronomical observations were ignored,
an alternative definition of ``fluid-frame luminosity'' of a jet could
be derived by noting the following: according to the world map of an
observer at rest in the fluid frame, the volume of the jet is
\emph{contracted}. Hence, $V\dpr=V/\Gamma$ and the fluid-frame
bolometric luminosity would be given by $L\dpr = 4\pi\,j\dpr V\dpr =
L/(\D\dpr^{3}\Gamma)$ --- i.e., exactly the opposite scaling of
$V\dpr$ and hence $L\dpr$ with $\Gamma$ from that given by
\citet[eqn.~A3]{SSO03}.  The reason for this apparent contradiction,
and that it is only apparently a contradiction, is again that result
of a relativistic experiment depends on what is held constant in which
frame (compare Appendix~\ref{s:illustr.lor}). All that matters for jet
observations is the supersnapshot with constant photon arrival times.

Equation~(\ref{eq:jetfluxdens}) of course correctly describes the
dependence of the received flux for a fixed jet feature when varying
the fluid's Lorentz factor and the line-of-sight angle.  As long as we
infer the correct observer-frame $V$ from the fixed apparent size of a
jet feature as function of the unknown fluid Lorentz factor and the
line-of-sight angle, we can infer the fluid-frame emissivity by
inverting that equation.  

For a spherical jet feature, the observer-frame volume can of course
be inferred directly from the projected radius $R$ of the sphere,
leading to
\begin{equation}
j\dpr_{\nu\dpr}(\nu) = S_\nu(\nu) \,\dL^{2}
\frac{\D\dpr^{-2+\alpha}}{4/3\pi R^3}.
\label{sphjet_jtrans}
\end{equation}

For a ``slab'' feature, we can measure its cross-sectional area $A$
and assume some rotational symmetry.  For fixed projected length
$\Lambda\proj$, the deprojected length is then just
$\Lambda\proj/\sin\theta$, so that the relation between emissivity and
observed flux of a jet section is
\begin{displaymath}
j\dpr_{\nu\dpr}(\nu) = S_\nu(\nu) \,\dL^{2}
\frac{\D\dpr^{-2+\alpha}\,\sin\theta}{A \Lambda\proj},
\end{displaymath}
again ignoring edge effects, i.e., assuming a length:width ratio of
greater than about 2.  This expression is identical to equation
(\ref{slabblob_jtrans}), the relation between observables and
rest-frame emissivity for a slab-shaped \emph{blob} of fixed projected
size but unknown $(\Gamma,\theta)$.  In other words, our lack of
knowledge about jet orientation and Lorentz factor affects elongated
slab-shaped blobs and jets in exactly the same way.

However, for a \emph{sphere} of fixed observed radius $R$, the
inferred fluid-frame emissivity is different by one power of
\D\ depending on whether we are considering a spherical blob or a
spherical jet feature (compare eqns.~\ref{sphjet_jtrans} and
\ref{sphblob_jtrans}).  This difference between the beaming of an
elongated and a spherical blob arises because a spherical blob in
effect becomes magnified \emph{along} the line of sight, while a slab
blob appears magnified \emph{in the plane of the sky}, as discussed in
\S\ref{s:beaming.blobs.restframe} above (again compare Figure
\ref{f:ray_sph}).  The elongation \emph{along} the line of sight is
unobservable in the projection on the plane of the sky. Thus, in the
special case of spherical features, there \emph{is} a difference
between ``jet'' and ``blob'' de-beaming.

\subsection{The minimum-energy magnetic field of beamed sources}
\label{s:beaming.Bmin}

A synchrotron source, such as the radio jet of an active galactic
nucleus, is powered by energy stored either in its magnetic field or
in relativistic particles. As first pointed out by \citet{Bur59},
there is a minimum to the total energy in particles and fields that is
required to power a given observed synchrotron luminosity.  This
minimum can be parameterized in terms of the magnetic field strength
at the minimum energy density, which is often used as ``the
minimum-energy magnetic field estimate''. 

Further derivations and ready-to-use formulae for non-relativistic
sources are given by \citet{Pach70}, \citet{Longair_vol2} and
\citet{Miley80}.  \citet{BK05} have offered some constructive
criticism on this so far well-established formalism, in particular
regarding the unknown ratio $K$ of total energy in relativistic
protons and electrons.  \citet{HK02} and \citet{SSO03} have presented
detailed derivations of minimum-energy estimates in the rest frame of
relativistic jets, but reached slightly different conclusions.  Armed
with the knowledge from the preceding sections, we can now reconsider
the question of the correct de-beaming of minimum-energy magnetic
field estimates.

As noted above, for the case of a ``jet'' geometry, the rest frame of
the fluid cannot be identified with the rest frame of the ``source'',
as the source volume is \emph{fixed} in the observer frame, while the
beaming is determined by the Lorentz factor at which the fluid is
\emph{moving} through the observer frame.  The same applies for
shocks. However, as extragalactic jets are assumed to be close to
ideal MHD conditions, the rest frame of the fluid is also the rest
frame of the magnetic field, and the frame in which the electron
energy distribution is assumed to be isotropic. Thus, whether we are
considering blobs or jets, the fluid rest frame is \Rev{the frame} in which
the minimum-energy field needs to be calculated \Rev{in order} to satisfy
the underlying assumptions. As before, the emission is assumed to be
optically thin emission with spectral shape
$j_{\nu\dpr}\propto\nu\dpr^\alpha$ and constant emissivity throughout
the emission region.

In the absence of beaming, the expression for the minimum-energy field
in terms of quantities in the rest frame of the fluid (double primes)
is
\begin{equation}
\label{eq:Bmin_dpr}
\Bmin\dpr^{7/2} \propto
\frac{\nu\dpr_2^{1/2+\alpha}-\nu\dpr_1^{1/2+\alpha}}{\nu\dpr_2^{1+\alpha}-\nu\dpr_1^{1+\alpha}}
\frac{L\dpr}{V\dpr},
\end{equation}
where the synchrotron spectrum extends over the rest-frame frequency
interval $(\nu\dpr_1,\nu\dpr_2)$; the first fraction is from the
function $\tilde c$ from \citet[eqn.\ 7.8]{Pach70}.  As pointed out
above, neither the rest-frame volume $V\dpr$ nor the observer-frame
volume $V=\D V\dpr$ is fixed by observations when the line-of-sight
angle is unknown, but it is the projection of the source on the plane
of the sky that needs to be kept fixed.  Nevertheless, it is
instructive to compare the minimum-energy field as function of
$\Gamma$ and $\theta$ inferred for fixed source volume to that
inferred for fixed projected appearance. \Rev{Hence both will be rederived
here, beginning with the fixed-source case
(\S\ref{s:disc.Bmin.volfixed}), followed by the fixed-projection case
\S\ref{s:disc.Bmin.obsfixed}. Their direct juxtaposition illustrates
why differing opinions have arisen in the literature about how
minimum-energy parameters scale with Doppler factor.}

\subsubsection{Minimum-energy parameters for fixed source volume}
\label{s:disc.Bmin.volfixed}

% \Rev{paragraph deleted}

\paragraph{Blob case} 
For a blob, the source volume is at rest in the fluid frame, therefore
$V\dpr$ is unambiguous, and the beamed minimum-energy field is
obtained from eq.~(\ref{eq:Bmin_dpr}) by inserting $L\dpr=\D\dpr^{-4}
L$, $V\dpr=\D\dpr^{-1}V$, and $\nu_{1,2}\dpr =
\D\dpr^{-1}\nu_{1,2}$. Hence, $\Bmin^{7/2}\propto \D\dpr^{-5/2}$ and
\begin{equation}
\label{eq:Bmin_Vfix_blob}
\Bmin(\D\dpr) = \Bmin(\D\dpr=1)\, \D\dpr^{-5/7}.
\end{equation}
This is identical to equation (A8) of \citet{SSO03}, which implies
that their expression applies to the fixed-volume case considered
here, not the fixed-observable case relevant to the interpretation of
observations.  It is different from equation (A7) of \citet{HK02}
because those authors explicitly considered the fixed-observable case
for a spherical blob whose rest-frame volume $V\dpr$ is inferred
directly from observables, the case that will be discussed in the
following subsection.

\paragraph{Jet case} 
From the frequency integral of the definition of monochromatic
luminosity (eq.~\ref{eq:Ldef}) and the $\D\dpr$-dependence of the
emissivity $j_{\nu}$ \citep[eq.~\protect\ref{eq:jtrans}; also eqn.\ 2.5
  of][]{LB85}, the bolometric luminosity of a jet section scales with
the Doppler factor of the fluid as $\D\dpr^3$.  Hence, $L\dpr/V$ for a
jet scales in exactly the same way as $L\dpr/V\dpr$ for a blob.  The
Doppler scaling of the frequencies occurring in the $\tilde c$
expression are identical to the blob case. Hence, the expression from
eq.~(\ref{eq:Bmin_Vfix_blob}) is obtained for the jet case also.
Again, this is the same result obtained by \citet{SSO03}, again
because they were implicitly considering the fixed-volume case.

\subsubsection{Minimum-energy field for fixed observables}
\label{s:disc.Bmin.obsfixed}

\Rev{This section describes how to infer the rest-frame minimum-energy
field from astronomical observations as function of the unknown
Lorentz factor and line-of-sight angle (usually only considered in
their combination as Doppler factor) in the case relevant to
observations, where it is not the source volume that is kept fixed,
but the source's projected appearance.  As in
\S\S\ref{s:beaming.blobs} and \ref{s:beaming.jets}, the computation of
the source volume needs to be done taking into account that changing
the Lorentz factor and line-of-sight angle not only changes the
Doppler factor, but also the \emph{length of the sight line through
  the object} due to the relativistic angle aberration
(eq.~\ref{eq:sintrans}). The differences between the formulae
presented below and those in the literature arise because the effects
of angle aberration are stated here explicitly.}

To express the minimum-energy field as function of an observed flux
density at some frequency $\nu\obs$, assuming a power-law spectrum,
one expresses the rest-frame flux density in terms of the rest-frame
luminosity
\begin{equation}
S\dpr_\nu(\nu\obs) = \frac{L\dpr}{4\pi \dL^2}
\frac{1+\alpha}{\nu\dpr_2^{1+\alpha}-\nu\dpr_1^{1+\alpha}}\,\nu\obs^{\alpha}.
\label{eq:Lobs_Snu}
\end{equation}
Substituting into eq.~(\ref{eq:Bmin_dpr}), one obtains
\begin{equation}
\Bmin\dpr^{7/2}  \propto  S\dpr_\nu(\nu\obs)\, \nu\obs^{-\alpha}\;
\frac{\nu\dpr_2^{1/2+\alpha}-\nu\dpr_1^{1/2+\alpha}}{V\dpr}.
\label{eq:Bmin_itoSnurest}
\end{equation}
As above (\S\ref{s:beaming.blobs}), the transformation properties for
fixed observables depend on the source geometry.

Before considering this expression for different source geometries, it
is important to note that the $\nu\obs$ term in this and the derived
expressions does \emph{not} scale with the Doppler factor, since it is
explicitly the fixed observing frequency.  The effect of observing the
rest-frame spectrum at a Doppler-shifted frequency and hence at a
different amplitude is already accounted for by the $-\alpha$
($K$-correction) term that appears in the exponent of the Doppler
factor multiplying the observed flux density. In other words, even
when $\Bmin$ is expressed in terms of $S_{\nu}(\nu\obs)$, which scales
as $\D\dpr^{3-\alpha}$, the spectral index $\alpha$ does not appear in
the exponent of $\D\dpr$ in the final expression for the
minimum-energy field, or the minimum energy content, because the
minimum-energy field depends on \emph{bolometric} quantities and all
$K$-correction terms drop out again.  This is also relevant for
deriving the boosted version of the yet-lower limit to the energy
content of a synchrotron source that \citet[p.\ 296,
  eqn.\ 19.29]{Longair_vol2} obtains by setting $\nu_1=\nu\obs$ in
eq.~(\ref{eq:Lobs_Snu}) and neglecting the other frequency
integration limit.  This substitution has to be done in the
Doppler-boosted expressions for the appropriate geometry, as derived
below. Hence, it would not be correct to set $\nu\dpr=\nu\obs$ in
eq.~(\ref{eq:Bmin_itoSnurest}) and apply the boosting afterwards; in
other words, the beamed version of eq.~(19.29) of
\citet{Longair_vol2} \emph{cannot} be obtained simply by inserting
Doppler factors into it. Once more, the correct boosting
transformation needs to be applied not only to the \emph{integrand},
but also to the \emph{integration boundaries}, and care has to be
taken to do both transformations at the same time.

\paragraph{Spherical blobs}

Since relativistically moving spheres are always observed with a
spherical outline, the rest-frame volume of a spherical blob can be
inferred directly from the observed radius $R$ as $V\dpr = 4/3\pi
R^3$.  Substituting this and the appropriate Doppler boosting for flux
density (eq.~\ref{eq:blobfluxdens}) and frequency
(eq.~\ref{eq:nutrans}) into eq.~(\ref{eq:Bmin_itoSnurest}) yields
\begin{equation}
\Bmin\dpr^{7/2} \propto  S_{\nu}(\nu\obs)\, \nu\obs^{-\alpha} \;
\frac{\nu_2^{1/2+\alpha}-\nu_1^{1/2+\alpha}}{4/3\pi R^3}\; \D\dpr^{-7/2}.
\label{eq:Bmin_sphblob}
\end{equation}
This implies that $\Bmin\dpr \propto \D\dpr^{-1}$, precisely eq.~(A7)
of \citet{HK02}, and reassuring since they explicitly considered only
this special case of a spherical blob.  However, hardly any of the
currently known jets has knots with morphology that accurately can be
described as spherical.

\paragraph{Slab or cylindrical blobs}

As discussed above, if a slab-shaped or cylindrical blob with
intrinsic length:width ratio of greater than 2 is observed to have a
projected length (image extent) $\Lambda\proj$ and cross-sectional
area $A$, its rest-frame volume is
$V\dpr=A\Lambda\proj/\D\dpr\,\sin\theta$.  Performing the
corresponding substitutions into eq.~(\ref{eq:Bmin_itoSnurest}), one
obtains
\begin{equation}
\Bmin\dpr^{7/2} \propto S_{\nu}(\nu\obs)\, \nu\obs^{-\alpha} \;
\frac{\nu_2^{1/2+\alpha}-\nu_1^{1/2+\alpha}}{A\,\Lambda\proj}\;
\D\dpr^{-5/2}\,\sin\theta.
\label{eq:Bmin_slabblob}
\end{equation}
The scaling of rest-frame minimum-energy field with the unknown
Lorentz factor and line-of-sight angle $\theta$ of such a blob is
$\Bmin\dpr \propto \D\dpr^{-5/7} (\sin\theta)^{2/7}$, i.e., it is not
expressible purely as function of the Doppler factor $\D\dpr$ because
the length of the sight line through the blob scales with
$\sin\theta$. The scaling differs from that of \citet{HK02} because
they considered only spherical blobs, and from that of \citet{SSO03}
because they did not consider the fixed-observable case.

\paragraph{Jet case}

In the jet case, the appropriate fluid-frame volume $V\dpr$ is again
the volume of fluid that contributes photons to the supersnapshot,
rather than the \emph{actual} Lorentz-contracted volume inferred from
the world map; hence, the appropriate volume is given by $V\dpr =
V/\D\dpr$.  For a jet with cross-sectional area $A$ and projected
length $\Lambda\proj$, the observer-frame volume is again
$V=A\Lambda\proj/\sin\theta$, and hence $V\dpr =
A\Lambda\proj/\D\dpr\,\sin\theta$ -- identical to the slab/cylindrical
blob case.  Therefore, eq.~(\ref{eq:Bmin_slabblob}) applies also to
continuous jets, analogous to the identical ``de-beaming'' equations
for slab blobs and jets in \S\S\ref{s:beaming.blobs} and
\ref{s:beaming.jets}.

\paragraph{An amusing special case}

Usually both $\theta$ and $\Gamma$ are unknown in jet
observations. However, the fact that the jet emission is detectable
makes it more likely that the emission is beamed towards the observer
than that it is beamed away from the observer.  A frequent guesstimate
that reduces the number of unknown beaming parameters from 2 to 1 is
therefore $\D=\Gamma$, equivalent to $\mu=\beta$ and
$\sin\theta=1/\Gamma$.  Substituting this assumption into
eq.~(\ref{eq:Bmin_slabblob}) leads to
\begin{displaymath}
\Bmin\dpr \propto 1/\Gamma = \sin\theta.
\end{displaymath}
Thus, if $\D=\Gamma$, the true fluid-frame minimum-energy field is
given by
\begin{eqnarray}
\Bmin\dpr &=& \Bmin^{(0)} \times 1/\Gamma \nonumber \\
 &=& \Bmin^{(0)} \times \sin\theta, \nonumber
\end{eqnarray}
where $\Bmin^{(0)}$ is the minimum-energy field inferred from the
observables by neglecting beaming and projection effects and assigning
the source a volume $A\Lambda\proj$.  It also turns out that at fixed
$\theta$, $\D\dpr^{-5/7} (\sin\theta)^{2/7} \geq 1/\Gamma$. Hence, the
true minimum-energy field at fixed $\theta$ is always larger than
$1/\Gamma \times \Bmin^{(0)}$.

\section{Discussion and summary}
\label{s:disc}

In \S\ref{s:appearance}, I have illustrated the difference between the
\emph{world map} of a relativistic jet, what is actually there, and
the \emph{world picture}, or its special case, the
\emph{supersnapshot}, that corresponds to what is observable by
distant astronomers.  For the quantitative interpretation of world
pictures in the presence of unknown beaming parameters (Lorentz factor
$\Gamma$ and angle $\theta$ between the fluid motion and the line of
sight), what matters is that the \emph{projected appearance} of the
jet is kept fixed and not the \emph{intrinsic volume}.  This gives
rise to de-beaming formulae that are slightly different from those in
the existing literature.

\subsection{Implications for interpretation of flux and surface
  brightness of jet features}

The most important conclusion for the quantitative analysis of jet
observations is that the scaling relations relating rest-frame
quantities (volume emissivity, intrinsic source size) to observables
(projected source size, surface brightness, total flux) \emph{cannot}
be stated as function of the line-of-sight angle and Lorentz factor in
a general way, but depend on the details of the source geometry.  It
is possible to write down explicit scaling relations for certain
simple geometries such as spheres and elongated, rotationally
symmetric blobs of constant cross-section. For other shapes, such as
ellipsoidal blobs or blobs with non-symmetric cross-sections, the
projected appearance is affected by edge effects, and additionally by
observational effects such as the contrast between the faintest parts
of the source and the sky background, as well as the available
signal-to-noise level.  Edge effects are properly taken into account
by the ray-tracing in \S\ref{s:illust.rays}, and such ray-tracing
modeling of observables is probably the most accurate route to
interpreting observations of relativistic objects. Indeed, it is part
of the prediction of observables from jet simulations such as those by
\citet{AMGea03}, e.g. \Rev{The work of \citet{SwiftHughes08}, which
explicitly considers the relation between the jet appearance in a
supersnasphot and the underlying physical quantities, taking into
account the retardation along the line of sight.}

\subsection{Implications for interpretation of morphological
  information in jet images}

Most radio, optical and X-ray maps of relativistic jets \citep[a list
  of radio jets is given by][]{LiuZhang02}\footnote{See
  \url{http://home.fnal.gov/~jester/optjets/} and
  \url{http://hea-www.harvard.edu/XJET/} for lists of optical and
  X-ray jets.} show a series of well-separated, distinct features
usually referred to as ``knots'', with diffuse emission linking them.
Given that relativistic beaming favours the detection of objects with
jets at small angles to the line of sight, and the superluminal
motions detected in the cores of many such sources, it is plausible
that the jet material itself is still relativistic even at large
separations (and indeed, this is required in models accounting for the
X-ray emission from powerful radio jets as beamed inverse-Compton
scattering of cosmic microwave background photons; see
\citealp{Tav00,Cel00} for the original development of the idea, as
well as the recent review by \citet{HK06}).  However, what is not
clear is whether the knots themselves are stationary shock features,
or themselves moving relativistically.

Referring to Figs.~\ref{f:ray_sph} and \ref{f:ray_slab}, the
prevalence of well-separated knots in jet images seems to suggest that
the knots are moving at least with mildly relativistic Lorentz factors
-- otherwise, there should be \Rev{\emph{some} jets observed at small
angles (favoured by Doppler boosting)} where different knots overlap
along the line of sight, washing out any individual morphological
features.

\Rev{In this case, the knots are subject to retardation magnification and
hiding (illustrated in Figs.~\ref{f:magnif} and \ref{f:angles}), and
we are not seeing \emph{all} of the jet features which are actually
present between core and hot spot, but just a small fraction (whose
magnitude is given by Fig.~\ref{f:visfrac}).  An observation of
apparent superluminal motion of individual knots would be a direct
confirmation that they are moving relativistically, as in the case of
parsec-scale knots in VLBI observations.  The kiloparsec-scale knots
are resolved out in VLBI observations, so that very long-term
monitoring programmes at sub-arcsecond spatial resolution are required
to make a potential superluminal motion observable.}

If jet knots are indeed moving relativistically, the retardation
magnification and hiding need to be taken into account when
interpreting morphological observables such as the ratio between knot
separation and jet width, which is important for addressing the
question of the origin of the jets' morphological features, e.g.,
whether they arise from instabilities \citep{Hardee03} or as
manifestation of a stable magnetohydrodynamical configuration
\citep{KC85}.

\subsection{Conclusion}

It becomes clear once more that relativistic effects are counter to
our non-relativistic intuition, and that familiarity with \emph{world
  maps}, Lorentz transformations and the resulting phenomena of length
contraction and time dilation is not sufficient for interpreting
\emph{world pictures}. When considering the beaming properties of
quantities expressed in terms of integrals, such as a surface
brightness, flux or the minimum-energy magnetic field estimate, one
needs to consider the transformation properties both of the integrand
and the integration volume.  Apparent differences between the beaming
properties of ``blobs'' and ``jets'' disappear when the same source
volume is considered.  Finally, the de-beaming of astronomical
observations needs to be done not for a fixed source, but for the
fixed \emph{projection} of the source.  Doing so resolves some
conflicts (again only apparent ones) between different de-beaming
formulae in the literature.  Given that astronomy provides only world
pictures, the concepts first laid out by \citet{Pen59} and
\citet{Terrell59} deserve more attention in the interpretation of jet
observations.

\section*{Acknowledgments}

This research has made extensive use of NASA's Astrophysics Data
System Bibliographic Services.  I am grateful to Herman Marshall for
valuable discussions and the impulse to begin this work, and
acknowledge fruitful interactions with members of the Fermilab
Experimental Astrophysics Group, and the Astronomy group at the
University of Southampton. I am particularly grateful to \L{}ukasz
Stawarz for detailed feedback in early stages of this work, to Jochen
Weller, Rob Fender and Dan Harris for continued discussions, to Arieh
K\"onigl for helpful comments on the paper -- in particular for
reminding me of the Lorentz invariance of opacity -- and to W.C.~Herren
for inspiration.  I thank the referee, Henric, Krawczynski, for his
constructive criticism which helped me improve the presentation of
this material. The portion of this work carried out at Fermilab was
supported through NASA contract NASGO4-5120A and through the
U.S.\ Department of Energy under contract No.\ DE-AC02-76CH03000.
Last but not least, I acknowledge support through an Otto Hahn
Fellowship from the Max-Planck-Institut f\"ur Astronomie that enabled
much of this work to be carried out within the Astronomy Group at the
University of Southampton, whose hospitality I enjoyed tremendously.

\appendix

\section{Lorentz expansion and time acceleration? A question of the
viewpoint}
\label{s:illustr.lor}

The world-map analysis  familiar from physics textbooks shows that
relativistically moving objects experience Lorentz contraction and
time dilation.  The derivation begins with the Lorentz transformations
between frames $\Sigma$ and $\Sigma\dpr$ with aligned $x$ and $x\dpr$
axes and origins coinciding at $t=t\dpr=0$:
\begin{eqnarray}
t &=& \gamma ( t\dpr + \beta x\dpr) \label{eq:Lor_t} \\
x &=& \gamma ( x\dpr + \beta t\dpr). \label{eq:Lor_x}
\end{eqnarray}
To show that there is Lorentz contraction of objects at rest in
$\Sigma\dpr$, one considers two observers at rest in $\Sigma$ which
coincide with opposite ends of the moving object at some fixed $t$,
e.g., $t=0$.  Solving eq.~(\ref{eq:Lor_t}) for $t\dpr$ and
substituting into eq.~(\ref{eq:Lor_x}) then yields $x_2-x_1 =
(x_2\dpr-x_1\dpr)/\gamma$, i.e., length contraction.  Time dilation is
obtained by considering a clock at rest in $\Sigma\dpr$ whose time is
read by two different observers in $\Sigma$, e.g., one at $(x,t) =
(0,0)$, the second at $(x,t)=(x_1,t_1=x_1/\beta)$.  The first observer
reads $t\dpr=0$ (by convention); setting $x\dpr = 0$ in
eq.~(\ref{eq:Lor_t}), the second observer reads $t\dpr = t/\gamma$
on the moving clock, i.e., infers that a shorter time interval has
elapsed in the moving frame than in the observers' frame and that time
is therefore dilated. 

However, we can ask our observers to do slightly different
experiments. Imagine that there are a large number of clocks in
$\Sigma\dpr$, and we are told that the clocks are all synchronized in
that frame. To investigate the behaviour of time, we ask a single
observer to compare the time read on successive clocks that are
whizzing past to her own observer-frame clock.  In other words,
instead of keeping $x\dpr$ fixed, let us keep $x$ fixed, and choose
the observer at $x=0$.  The first clock reads $t\dpr=0$.  If the
clocks are separated in $\Sigma\dpr$ by some distance $l\dpr$, the
second clock, being at $x\dpr=-l\dpr$ will reach the observer at
$t\dpr = l\dpr/\beta$.  At this time, the observer's own clock reads
$t=l\dpr/(\beta\gamma) = t\dpr/\gamma$; thus, the observer's clock has
advanced less than the moving ones, and time is accelerated instead of
dilated. The reconciliation with time dilation is, of course, that the
second moving clock did not read $t\dpr=0$ at $t=0$, but already
$t\dpr=\beta l\dpr$ --- the clocks that are synchronized in
$\Sigma\dpr$ are \emph{not} in $\Sigma$, simultaneity is relative.
Indeed, this setup allows the observers at rest in $\Sigma$ to infer
that \emph{their} clocks are slowed down relative to observers in
$\Sigma\dpr$: it compares the time interval elapsed on a \emph{single}
clock in one frame to the interval elapsed between two
\emph{different} clocks in another frame.  The individual clock runs
slow compared to two different clocks whizzing past it, or that it
whizzes past.

We can also derive an alternative length measurement.  Assume that the
observers know about special relativity, and that the observers in
$\Sigma\dpr$ have placed a clock at the front and rear end of the
object whose length the observers in $\Sigma$ are trying to measure.
These observers decide to measure the length by taking into account
the relativity of simultaneity, and they do so by noting the position
of each clock in their own frame when it shows a fixed time in the
\emph{moving} frame, $t\dpr=0$, say.  If the rear clock is at
$x\dpr=0$, it reads $t\dpr=0$ at $t=0$ and will be seen by the
observer at $x=0$.  The front clock is at $x\dpr=l\dpr$, and thus from
eq.~(\ref{eq:Lor_x}) it will read $t\dpr=0$ when it has reached the
observer at $x=\gamma l\dpr$.  Hence, the observers in $\Sigma$ infer
that the moving rod has expanded by a factor $\gamma$ compared to what
it is in its own rest frame.  Again, this is just the converse of the
observers in $\Sigma\dpr$ measuring the length of an object at rest in
$\Sigma$ and finding that the moving object has contracted.

Thus, observers in either frame can both observe that moving objects
appear contracted to them \emph{and} infer that something that is at
rest in their own frame will appear contracted to observers in the
other frame; similarly, observers can observe clocks in the other
frame run slow \emph{and} infer that their own clocks will be observed
to run slow by observers in the other frame. The important point is
that the outcome of an experiment depends on \emph{which quantity is
  held fixed in which frame}, and the correct interpretation of
apparently contradictory results depends on being clear in the
description of the experiment.

%\bibliographystyle{mn2e}
%\bibliography{mn-jour,jester_master}
% \input{volumetrans.bbl}
% BEGIN INPUT from volumetrans.bbl

% END INPUT from volumetrans.bbl

\end{document}